\documentclass[a4paper,11pt]{article}
\pdfoutput=1 % if your are submitting a pdflatex (i.e. if you have
\usepackage{jheppub} % for details on the use of the package, please
\usepackage{slashed}

\usepackage[T1]{fontenc} % if needed

\title{\boldmath Nonrelativistic strings and the limits of $ \mathcal{N}=2 $ dualities}

\author[a]{Dibakar Roychowdhury}
\affiliation[a]{Department of Physics, Indian Institute of Technology Roorkee,\\Roorkee 247667, Uttarakhand, India}
\emailAdd{dibakar.roychowdhury@ph.iitr.ac.in}
\abstract{We carry out the formulation of nonrelativistic (NR) F-string theory for $ \mathcal{N}=2 $ Gaiotto-Maldacena class of geometries in type IIA supergravity. Considering the NS-NS sector, we reduce the type IIA theory into the NR sigma models those are formulated over the nonrelativistic target space geometries known as the torsional String Newton-Cartan (TSNC) backgrounds. Using these TSNC data, we write down the sigma model action and explore several properties in the semiclassical limit. In particular, we carry out the \emph{canonical} analysis of the T-duality transformation rules in TSNC background. Following the standard canonical prescription of obtaining T-duality transformations, we map these TSNC sigma models to relativistic sigma models propagating over general relativity backgrounds. Finally, we conclude with a detailed discussion on the effects of adding \emph{flavour} branes and thereby taking the corresponding TSNC limit in the bulk description of $ \mathcal{N}=2 $ quivers. We compute various field theory observables in the nonrelativistic limit and outline the structure of the NR Lagrangian following the light cone reduction of $ \mathcal{N}=2 $  SYM theory.}
\begin{document} 
\maketitle
\flushbottom
\newpage
%%%%%%%%%%%%%%%%%%%%%%%%%%%%%%%%%%%%%%%%%%%%%%%%%%%%%%%%%%
\section{Introduction and summary}
Nonrelativistic (NR) string theory, as was originally formulated for relativistic strings propagating on flat space \cite{Gomis:2000bd} has gained renewed attention in the recent years both in the context of NR holography and string theory \cite{Gomis:2005pg}-\cite{Harmark:2019upf}. The inclusion of NS-NS fluxes, however, has remained as a challenge for decades and requires the addition of new symmetry generators/global charges in the relativistic theory leading towards what is known as the string Poincare algebra. In the NR counterpart, one takes the large $ c(\rightarrow \infty) $ limit of string Poincare algebra which eventually leads to F- string Galilei algebra \cite{Bidussi:2021ujm}.

One of the prime motivations of the present work is to study T-duality rules for NR closed string sector with background NS-NS fluxes. T-duality for nonrelativistic (NR) closed strings in String Newton-Cartan (SNC) background was first investigated by authors in \cite{Bergshoeff:2018yvt} which was further generalised for open strings in \cite{Gomis:2020izd}. The present paper is an attempt to generalise these results in the presence of NS-NS sector where the natural interpretation of the target space geometry appears to be the torsional String Newton-Cartan (TSNC) background \cite{Bidussi:2021ujm}-\cite{Yan:2021lbe} rather than the ordinary String Newton-Cartan (SNC) background.

We follow the standard \emph{canonical} approach \cite{Alvarez:1994wj}-\cite{Alvarez:1994dn} of obtaining the T-duality transformation rules. As a concrete application of their formalism, we take the specific example of TSNC limit of non-Abelian T duals (NATD) of $AdS_5 \times S^5$ and apply the canonical methods to generate new classes of \emph{relativistic} sigma models those are propagating over general relativity backgrounds. These results therefore reestablish the previous claim of \cite{Bergshoeff:2018yvt} in the presence of background NS-NS fluxes. 

The general relativity NATD backgrounds fall under the type IIA category and appear with a natural NS-NS sector while preserving the background $\mathcal{N}=2$ supersymmetry \cite{Gaiotto:2009we}-\cite{Gaiotto:2009gz}. These NATD geometries can be classified into two categories - (i) the Sfetsos-Thompson (ST) background \cite{Sfetsos:2010uq}-\cite{Maldacena:2000mw} and its flavour generalisations \cite{ReidEdwards:2010qs}-\cite{Aharony:2012tz} those are dual to $\mathcal{N}=2$ linear quivers in $ 4d $ \cite{Lozano:2016kum}-\cite{Nunez:2019gbg} and (ii) the NATD of $ AdS $ \cite{Lozano:2017ole} those are dual to ($ 0+1 $)d BMN matrix model vacua \cite{Berenstein:2002jq}.

The purpose of the present paper is to take a systematic large $ c $ limit of these $ \mathcal{N}=2 $ geometries and write down the corresponding NR sigma model in the TSNC limit \cite{Bidussi:2021ujm}. These NR sigma models serve as a platform to explore a number of interesting properties (which we list below) including the T-duality in the framework of NR holography. 
%%%%%%%%%%%%%%%%%%%%%%%%%%%%%%%%%
\subsection{TSNC limit and NR strings:A quick review} 
 For the familiarity of the reader, below we summarise the basic aspects of TSNC limit and the associated algebra. The discussion below is a summary of work done in \cite{Bidussi:2021ujm}.

The inclusion of the $ B $ field (together with the metric) is taken care of by introducing a new set of gauge fields ($ \pi_{\mu}^{\hat{a}} $) along with their respective generators $ Q_{\hat{a}} $.  Here, $ \hat{a}=0, \cdots , d-1 $ stands for the spacetime indices of the $ d $ dimensional general relativity background.

This defines the following gauge connection
\begin{eqnarray}
\mathcal{A}_{\mu}=e_{\mu}^{\hat{a}}P_{\hat{a}}+\frac{1}{2}\omega_{\mu}^{\hat{a}\hat{b}}M_{\hat{a}\hat{b}}+\pi_{\mu}^{\hat{a}}Q_{\hat{a}},
\end{eqnarray}
on the extended Poincare algebra known as the string Poincare algebra. Clearly, the gauging procedure of the string Poincare algebra using the complete set of data $ \lbrace e_{\mu}^{\hat{a}} , \omega_{\mu}^{\hat{a}\hat{b}} , \pi_{\mu}^{\hat{a}} \rbrace$ defines the target space geometry for the NS-NS sector of the relativistic closed string theory.

While taking the NR ($ c \rightarrow \infty $) limit, the spacetime indices are decomposed into longitudinal ($ A=0,1 $) as well as transverse indices ($ a=2, \cdots , d-1 $). Following this decompositions\footnote{The details of these decompositions have been discussed in section \ref{2.2}.} of the frame fields as well as the $ \pi $ fields which when substituted into the relativistic string action result into the following entities
\begin{eqnarray}
\tau_{\mu}^A ~;~h_{\mu \nu}=e_{\mu}^{a}e_{\nu}^b \delta_{ab}~;~B^{(NR)}_{\mu \nu}=\frac{1}{2}\delta_{ab}(e_{\mu}^a \pi_{\nu}^b -e_{\nu}^a \pi_{\mu}^b)~;~m_{\mu \nu}=\eta_{AB}\tau_{[\mu}^{A}\pi_{\nu]}^B +B^{(NR)}_{\mu \nu}.
\end{eqnarray} 

Here, $ \tau_{\mu}^A $ is the clock one form, $ h_{\mu \nu} $ is the transverse metric and $ B^{(NR)}_{\mu \nu} $ is the transverse magnetic field. The set $ \lbrace \tau_{\mu}^A , h_{\mu \nu} , m_{\mu \nu} \rbrace$ together constitutes what we define as the TSNC data. Therefore, given a general relativity background, the first step to construct the corresponding NR sigma model is to write down the frame fields and the $ \pi $ fields in the large $ c $ limit to extract the above TSNC data.
%%%%%%%%%%%%%%%%%%%%%%%%%%%%%%%%%
\subsection{Summary of results} 
Following the spirit as mentioned above, we obtain the parent NR sigma model taking a large $c$ limit of $\mathcal{N}=2$ backgrounds. We collect the TSNC data and identify the corresponding target space geometry as the TSNC limit \cite{Yan:2021lbe}-\cite{Bidussi:2021ujm} of $\mathcal{N}=2$ backgrounds. These TSNC data were further analysed to explore a number of properties of the NR sigma model.

These TSNC data are what one should be able to recast as a solution of the type IIA supergravity equations in the TSNC limit. TSNC limit of the supergravity equations of motion can be obtained starting from a (bosonic) action in the Einstein frame \cite{Henneaux:2008nr}
\begin{eqnarray}
\label{EEE1.3}
S \sim \int d^{10}x\sqrt{-\mathcal{G}}(\mathcal{R}-\frac{1}{2}| \partial \phi |^2 -\frac{1}{12}e^{-\phi}| \mathcal{H}_3 |^2 + \cdots),
\end{eqnarray}
and thereby decomposing the metric and the B field into longitudinal ($ A=0,1 $) and transverse ($ a=2, \cdots ,8 $) components as \cite{Bidussi:2021ujm}
\begin{eqnarray}
\label{EEE1.4}
\mathcal{G}_{\mu \nu}=c^2 E^A_{\mu}E^B_{\nu}\eta_{AB}+\delta_{ab}e^a_{\mu}e^b_{\nu}~;~B_{\mu \nu}=c^2 \eta_{AB}E^A_{[\mu}\Pi^B_{\nu]}+\delta_{ab}e^a_{[ \mu}\pi^b_{\nu]},
\end{eqnarray}
where $ E^A_{\mu} $ are the longitudinal vielbein those can be expressed using the TSNC data (\ref{e11}). 

On a similar note, $ \Pi^a_{\mu} $ fields can also be expanded using TSNC data \cite{Bidussi:2021ujm}
\begin{eqnarray}
\Pi^a_{\mu} = \epsilon^A_B \tau_{\mu}^B +\frac{1}{2c^2}\pi_{\mu}^A.
\end{eqnarray}

Upon substituting (\ref{EEE1.4}) into (\ref{EEE1.3}) and thereby taking a large $ c (\rightarrow \infty )$ limit, the resulting nonrelativistic equations of motion should be read off from the leading order \emph{finite} action. It would be really nice to pursue these equations further and investigate nonrelativistic limits of type IIA supergravity dynamics\footnote{For a similar analysis, the reader is reffered to \cite{Bergshoeff:2021bmc}. }. 

In order to discuss the effects of adding flavour branes, we adopt the change in coordinates \cite{Lozano:2016kum} and rewrite the ST using a single potential function $ V(\sigma , \eta ) $ that satisfies Laplace's equation of electrostatics \cite{Gaiotto:2009gz}. The solution is further characterised by a single charge density $ \lambda (\eta) $ which allows us to classify $ \mathcal{N}=2 $ quivers into different categories\footnote{See \ref{sk} and \ref{ul} for details.} \cite{ReidEdwards:2010qs}. 

To start with, we expand the relativistic target space near $ \sigma \sim 0 $ and thereby taking its TSNC limit which results in a new subclasses of NR backgrounds for which the sigma model is defined. For the purpose of this paper, we consider $ \lbrace t , \eta \rbrace $ to be the longitudinal directions while the coordinates $ \lbrace \chi , \xi\rbrace $ along the two sphere are taken to be transverse.

\paragraph{Nonrelativistic string spectra and Integrability.} The TSNC limits of NATD geometries serve as a profound platform to explore NR string spectra using the \emph{semiclassical} approximations and in fact going beyond it. As our analysis reveals, one can estimate the spectra of NR spinning strings going beyond semiclassical approximations in the NR t'Hooft coupling ($ \lambda_{NR} $) which shows, $ \mathcal{E}_{NR}\sim \mathcal{E}_{NR}^{(cl)}+\mathcal{O}(1/\sqrt{\lambda_{NR}}) $ and $ \mathcal{S}\sim \mathcal{S}^{(cl)}+\mathcal{O}(1/\lambda^{1/4}_{NR}) $. A similar analysis in the TSNC limit of ST background reveals NR rotating string spectra with energy $ \mathcal{E}_{NR}\sim \mathcal{E}_{NR}^{(cl)}+\mathcal{O}(1/\lambda_{NR} ) $ and the R charges those behave as $ \mathcal{J}_{1,2}\sim\mathcal{J}^{(cl)}_{1,2} +\mathcal{O}(1/\lambda^{1/4}_{NR}) $.

The next question that arises is whether these semiclassical strings are integrable or not. To address this issue, we further explore the Liouville integrability of the canonical phase space associated with these NR strings. This is based on the algorithm\footnote{See the section \ref{liouville} for a detailed discussion on the algorithm.} provided due to Kovacic in \cite{kova} which was implemented for strings \cite{Basu:2011fw} as well as branes \cite{Stepanchuk:2012xi} with a subsequent application to (relativistic) ST backgrounds \cite{Nunez:2018qcj} and NR sigma models \cite{Roychowdhury:2019olt}.

Like in the relativistic example \cite{Nunez:2018qcj}, the TSNC limit of NATD background preserves the Liouville integrability criteria in the sense of Kovacic \cite{kova}. We verify this issue separately both for the TSNC limit of NATD $ AdS $ and the ST background. This ensures the classical integrability of these NR sigma models in the TSNC limit of NATD $ AdS_5 \times S^5 $. 

However, as we see, the addition of flavour D6 branes into the picture reveals something interesting and in fact is quite reminiscent of its relativistic counterpart \cite{Nunez:2018qcj}. It turns out that it is the presence of the flavour D6 branes in the parent NATD theory that forbids the NR strings from being integrable while taking the TSNC limit\footnote{see sections \ref{flavb1} and \ref{flavb2} for a detailed analysis and results.}. To summarize, the TSNC limit of ST turns out to be a special limiting case of $ \mathcal{N}=2 $ solutions which preserves the classical integrability of NR strings in the sense of Kovacic \cite{kova}. 

\paragraph{T-duality and QFT observables.} We carry out a general analysis of T-duality transformation rules following the canonical approach of \cite{Alvarez:1994wj}-\cite{Alvarez:1994dn}. Considering the NR target space geometry as the TSNC limit of the NATD $AdS_5 \times S^5$, the T-duality is performed along (one of the) the transverse isometry directions of the non Lorentzian background.  During the process, we land up in a new class of relativistic sigma models propagating over (pseudo) Riemannian manifolds. 

Below, we summarise theses T-duality transformation rules for both classes of NR sigma models. For NR spinning strings (\ref{e2.26}) (those are propagating) in NATD $ AdS $ these rules are summarised as: 
\begin{eqnarray}
\tilde{g}_{\eta \eta}&=&g_{\eta \eta}=\frac{1}{\varrho^2}~;~\tilde{g}_{\varrho \varrho}=\frac{\ell^4}{4 \alpha'^2_{NR}}\varrho^2 ,\\
\tilde{g}_{\chi \chi}&=& g_{\chi \chi}+\frac{\ell^4}{16 \alpha'^4_{NR}}\frac{B^{2(NR)}_{\chi \xi}}{g_{\xi \xi}},\\
\tilde{g}_{\chi \tilde{\xi}}&=&\frac{\ell^2}{4 \alpha'^2_{NR}}\frac{B^{(NR)}_{\chi \xi}}{g_{\xi \xi}}~;~\tilde{g}_{\tilde{\xi}\tilde{\xi}}=\frac{1}{4g_{\xi \xi}},
\end{eqnarray}
where $ \tilde{g}_{\mu \nu} $ are the metrics of the general relativity background.

Here, $ \lbrace \eta , \chi \rbrace $ are the transverse coordinates \cite{Bidussi:2021ujm} of the TSNC target space manifold while on the other hand, $ \xi \in S^2 $ is the transverse isometry direction along which the T-duality is applied. We also identify, $ B^{(NR)}_{\chi \xi} (\sim \frac{ \alpha'^3_{NR}\eta^3 \sin\chi}{\ell^4 \varrho^4} )$ as the NS-NS flux associated with the non Lorentzian manifold in the TSNC limit of NATD $ AdS $.

\begin{figure}
\includegraphics[scale=.65]{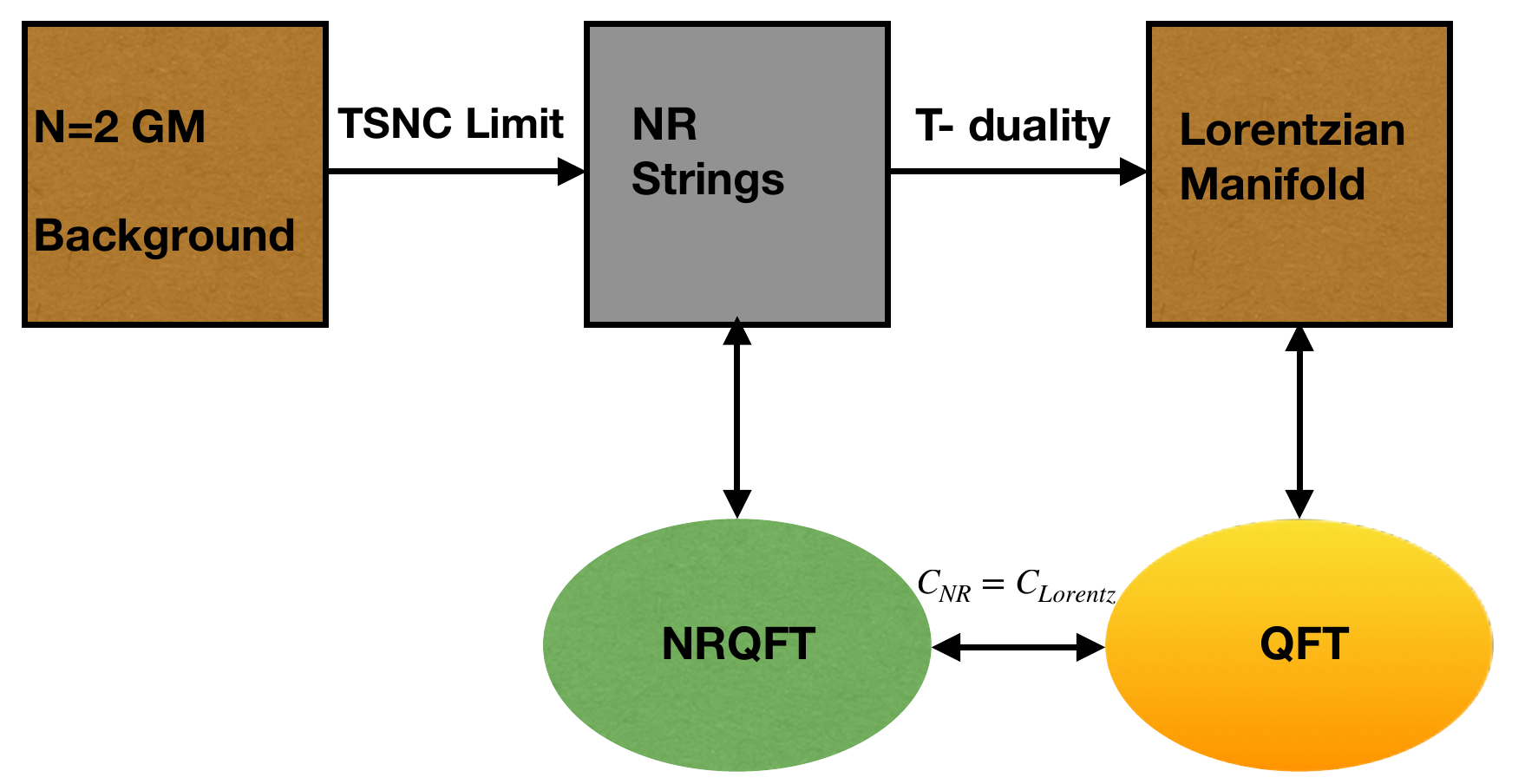}
  \caption{The above flow chart summarises the key idea of the paper. The Lorentzian background is obtained following a T-duality along the transverse isometry circle of the non-Lorentzian manifold. As a consequence of this, the stringy descriptions in these backgrounds are equivalent and compute the unique central charge for the dual QFTs.} \label{fchart}
\end{figure}

A similar analysis for rotating strings (\ref{E2.60}) in the TSNC limit of ST background reveals an identical set of T-duality rules:
\begin{eqnarray}
\label{natd 1}
\tilde{g}_{ab}&=&g_{ab}~;~a,b = \beta , \psi \\
\tilde{g}_{\theta \theta}&=& g_{\theta \theta}+\frac{\ell^4}{16 \alpha'^4_{NR}}\frac{B^{2(NR)}_{\theta \varphi}}{g_{\varphi \varphi}},\\
\tilde{g}_{\tilde{\varphi}\theta}&=&\frac{\ell^2}{4\alpha'^2_{NR}}\frac{B^{(NR)}_{\theta \varphi }}{g_{\varphi \varphi}}~;~\tilde{g}_{\tilde{\varphi}\tilde{\varphi}} = \frac{1}{4g_{\varphi \varphi}},
\label{natd2}
\end{eqnarray}
which correspond to relativistic rotating strings propagating over Lorentzian manifolds.

Here, $ \lbrace \beta , \psi \rbrace $ are the transverse coordinates of the NR target space. On the other hand, $ \lbrace \theta , \varphi \rbrace \in S^2$ of the TSNC limit of ST background - where in particular, we identify the circle $ 0 \leq \varphi \leq 2\pi $ as being the direction of isometry along which the T-duality is applied.

While adding flavour D6 branes, our analysis reveals identical TSNC limits for both \emph{single kink} and the \emph{Uluru} profiles. This translates into the fact that the T-duality rules for both of these configurations are identical\footnote{See section \ref{3.2.2} for details.} to that of (\ref{natd 1})-(\ref{natd2}).

The relativistic background that is found to be T dual to NR strings provides an excellent platform to carry out further analysis on various field theory observables at strong coupling and shed light on its dual NR counterpart. We compute various field theory observable, for example the central charge \cite{Macpherson:2014eza} which is therefore conjectured to be the same as that of the parent NR theory since they are related by a transverse T-duality.

Below, we summarize our key findings on the central charge corresponding to different general relativity configurations. These are precisely the central charges of the parent NR theories those are obtained as a TSNC limit of $ \mathcal{N}=2 $ backgrounds (Fig. \ref{fchart}),
\begin{eqnarray}
\label{central}
c_{NR} \sim \begin{cases}
      N^3_5 +\mathcal{O}(N^{-1}_5)&  (\text{NR Single Kink})\\
      N^2_5 +\mathcal{O}(N^{-1}_5)& (\text{NR Uluru})
    \end{cases} 
\end{eqnarray}
where $ N_5 $ is the number of NS5 branes associated with NATD backgrounds.

As our analysis reveals, the central charge corresponding to the TSNC limit of ST backgrounds goes as $ \sim N^2_5 $ which turns out to be infinite as there are no upper bounds on the number of NS5 branes for ST. On the other hand, the expressions in (\ref{central}) are finite as the number of NS5 branes for the parent NATD backgrounds are bounded due to the presence of the flavour D6 branes.

\begin{figure}
\includegraphics[scale=.70]{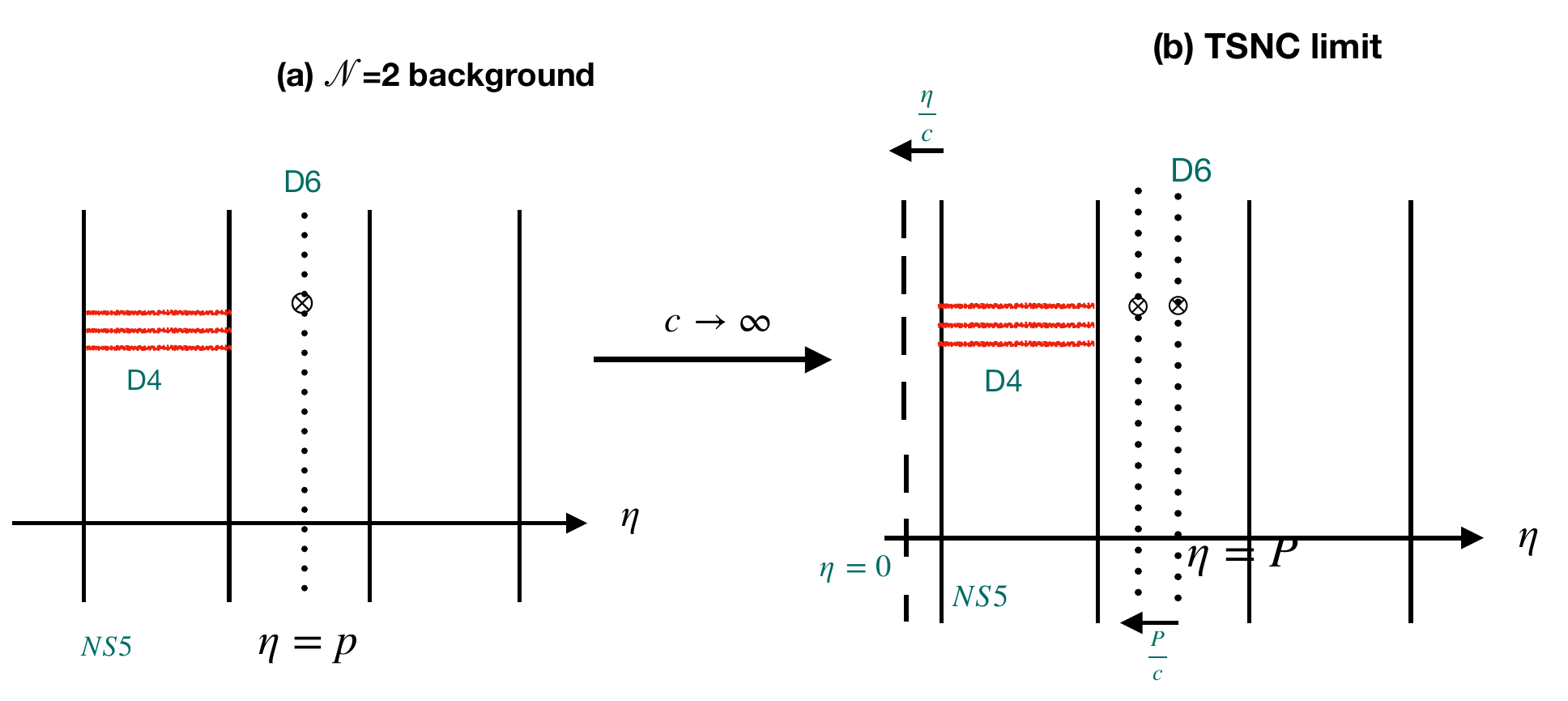}
  \caption{The above figure elaborates the brane set up of the $ \mathcal{N}=2 $ Hanany-Witten (HW) configuration in the large $ c $ limit. The left figure (a) represents the original HW set up of the Gaiotto-Maldacena background. The right figure (b) shows the shifted positions of the NS5 branes as well as the flavour D6 ranes under NR scaling. In the strict NR ($ c \rightarrow \infty $) limit all the NS5 branes collapse at the origin of the holographic ($ \eta $) axis making the theory strongly coupled. We show this explicitly in section \ref{conclud} by computing the coupling in the NR limit.} \label{hw}
\end{figure}

As our analysis shows, the NR scaling eventually shifts ($ \eta \rightarrow \frac{\eta}{c} $) the previous locations of NS5 as well as flavour D6 branes more towards the origin along the holographic $ \eta $ axis. It turns out that in the TSNC limit, some of these NS5 branes are eventually placed at the origin $ \eta \sim 0 $ causing a singularity in the metric of the NR target spacetime (see Fig. \ref{hw}). 

As a result of such NR scaling, one has  to put a lower cut-off ($ \eta_{min} $) to avoid singularities in the central charge calculations. A similar line of arguments defines a modified upper bound $ \eta_{max}\sim \frac{N_5}{c} $ along the (rescaled) holographic axis of the NR target spacetime.

The rest of the paper deals with generalising the previous results for generic Gaiotto-Maldacena class of geometries \cite{Gaiotto:2009gz} (see section \ref{GM}). Taking a number of examples, we show that these generalised formulae boil down into desired expressions in the large $ c \rightarrow \infty $ limit. 

Finally, we conclude in section \ref{conclud}, where we provide further insights into the NR limit of the $ \mathcal{N}=2 $ SCFTs by estimating some more QFT observables. These include the couplings of the NR theory and an outline on the structure of the NR Lagrangian.
%%%%%%%%%%%%%%%%%%%%%%%%%%%%%%%%%%%%%%%%
\section{Nonrelativistic limits without flavour}
%%%%%%%%%%%%%%%%%%%%%%%%%%%%%%%%%%%%%%%%
\subsection{$  \mathcal{N}=2$ backgrounds without flavour}
We begin by summarising the full 10d type IIA background that is relevant for our analysis
\begin{eqnarray}
\label{e1}
ds_{10}^2 = ds^2_{NATD}+ds^2_{ST}.
\end{eqnarray}

The NATD $ AdS $ is obtained by applying T-duality along $ SU(2) $ of $ AdS_5 $ \cite{Lozano:2017ole}
\begin{eqnarray}
\label{e2}
ds^2_{NATD}=L^2(-\cosh^2 \varrho dt^2 +d\varrho^2)+\frac{4 \alpha'^2 d\eta^2}{L^2 \sinh^2 \varrho} \nonumber\\
+\frac{4L^2 \alpha'^2 \eta^2 \sinh^2\varrho}{(16 \alpha'^2 \eta^2 +L^4 \sinh^4\varrho)}(d\chi^2 +\sin^2\chi d\xi^2),
\end{eqnarray}
where $ \varrho_{min}\leq\varrho \leq \infty $ stands for the radial coordinate in global coordinates. 

The other (global) coordinate $ \eta \sim N_5 $, on the other hand, has an interpretation in terms of the rank of the irreducible representation in the partition of $ N $, where $ N $ stands for the rank of the $ SU(2) $ gauge group in the dual BMN matrix model.

The spacetime singularity in the metric at $ \varrho \sim 0 $ is an artefact of the presence of NS5 branes. These geometries fall under the general half-BPS category of Lin and Maldacena (LM) which are dual to non BMN vacuum \cite{Lozano:2017ole} of Plane Wave Matrix Model (PWMM).

In a similar manner, the NATD  $ S^5 $ is obtained by applying T-duality transformations\footnote{Also known as the Sfetsos-Thompson (ST) solution \cite{Sfetsos:2010uq}.} \cite{Sfetsos:2010uq} along $ SU(2)\subset SO(6) $ which results in the metric of the following form
\begin{eqnarray}
\label{e3}
ds^2_{ST}=L^2(d\alpha^2 + \sin^2\alpha d\beta^2)+\frac{4\alpha'^2 d\psi^2}{L^2 \cos^2\alpha}\nonumber\\
+\frac{4L^2 \alpha'^2 \psi^2 \cos^2\alpha}{(16 \alpha'^2 \psi^2 +L^4 \cos^4\alpha)}(d\theta^2 + \sin^2\theta d\varphi^2),
\end{eqnarray}
that is conjectured to be dual to $ \mathcal{N}=2 $ linear quivers \cite{Lozano:2016kum} in $ 4d $.

The type IIA geometry (\ref{e2})-(\ref{e3}) is supplemented by the background NS-NS and RR fluxes of the following form\footnote{For the purpose of our present paper, we focus only on the NS-NS sector of the closed string dynamics. The primary reason for this is that the associated TSNC data  and the corresponding generators of the F- string Galilei algebra are clearly worked out \cite{Bidussi:2021ujm}. Our strategy would be to introduce NR scaling (\ref{e12}) and (\ref{E2.50}) to extract these geometric data in the large $ c $ limits of the metric and the B field \cite{Bidussi:2021ujm}. We use these TSNC data to writedown the corresponding TSNC sigma model action (\ref{e22}). Likewise, the supersymmetric generalisation of TSNC sigma models (\ref{e22}) require insights into the \emph{geometric} realisation of the RR fields and thereby a supersymmetric extension of the string Poincare algebra \cite{Bidussi:2021ujm}. While taking the TSNC limit, this should generate the TSNC data for the RR sector which would play a pivotal role in building up a supersymmetric generalisation of F-string Galilei algebra \cite{Bidussi:2021ujm}. Similar remarks also hold for the dilaton field of the NS-NS sector which would again require a nontrivial extension of the string Poincare algebra and subsequently its nonrelativistic limit. Like in the NS-NS sector, one could naively consider a nonrelativistic limit of the RR sector following the NR scaling (\ref{e12}) and (\ref{E2.50}). However, unlike the NS-NS sector, the geometric data corresponding to the RR sector are not settled down yet. In other words, for the RR sector, it is a priori not clear what kind of TSNC data the above nonrelativistic scaling (\ref{e12}) and (\ref{E2.50}) would correspond to. We leave these issues for future investigations. }
\begin{eqnarray}
-B_2 &=& \frac{16 \alpha'^3 \eta^3}{(16 \alpha'^2 \eta^2 +L^4 \sinh^4\varrho)}\sin\chi d\chi \wedge d\xi 
+\frac{16 \alpha'^3 \psi^3}{(16 \alpha'^2 \psi^2 +L^4 \cos^4\alpha)}\sin\theta d\theta \wedge d\varphi , \\
F_2 &=& \frac{L^4}{2 \alpha'^{3/2}}\cosh\varrho \sinh^3\varrho d\varrho \wedge dt +\frac{L^4}{2 \alpha'^{3/2}}\cos^3\alpha \sin\alpha d\alpha \wedge d \beta , \\
F_4 &=& B_2 \wedge F_2.
\end{eqnarray}

Finally, we note down the solution for the background dilaton
\begin{eqnarray}
e^{-2\phi}=\frac{L^2 \sinh^2\varrho}{64 \alpha'^3}(16 \alpha'^2 \eta^2 +L^4 \sinh^4\varrho)+\frac{L^2 \cos^2\alpha}{64 \alpha'^3}(16 \alpha'^2 \psi^2 +L^4 \cos^4\alpha).
\end{eqnarray}
%%%%%%%%%%%%%%%%%%%%%%%%%%%%%%%%%%%%%%%%%%%%%%%%%%%%%%%%%%%%
\subsection{The BMN vacua}
\label{2.2}
Before we get into the discussion on Gaiotto-Maldacena (GM) backgrounds, we would like to comment on the NR limit of BMN vacua \cite{Berenstein:2002jq} which are dual to NR strings those are propagating over non-Lorentzian manifolds obtained as the TSNC limit of NATD $ AdS $. To start with, we switch off the coordinates of the internal manifold (\ref{e3}) and focus only on the dynamics in NATD $ AdS $. In order to map NATD $ AdS $ geometries to torsional string Newton-Cartan (TSNC) type backgrounds we follow the prescriptions of \cite{Gomis:2005pg}, \cite{Bidussi:2021ujm}. 

We introduce vielbein fields $ e^{\hat{a}}_{\mu} $ such that the target space metric can be expressed as
\begin{eqnarray}
\mathcal{G}_{\mu \nu}= e^{\hat{a}}_{\mu}  e^{\hat{b}}_{\nu} \eta_{\hat{a}\hat{b}},
\end{eqnarray}
where $ \hat{a}, \hat{b} $ are the tangent space indices such that $ \eta_{\hat{a}\hat{b}}=\text{diag}(-1,1,\cdots,1) $.

Following \cite{Bidussi:2021ujm}, we further decompose the tangent space indices into longitudinal ($ A=0,1 $) and transverse ($ a=2,\cdots, 9 $) directions such that \cite{Bidussi:2021ujm}
\begin{eqnarray}
e^{\hat{a}}_{\mu}=(c E^{A}_{\mu}, e^{a}_{\mu})~;~\pi^{\hat{a}}_{\mu}=(c \Pi_{\mu}^{A}, \pi_{\mu}^a)
\end{eqnarray}
where $ \pi^{\hat{a}}_{\mu} $ are the gauge fields \cite{Bidussi:2021ujm} corresponding to background NS-NS fluxes.

In what follows, we obtain the corresponding nonrelativistic (NR) version of these semiclassical strings which we denote as type IIA torsional string Newton-Cartan limit. In other words, the world-sheet vielbeins that we evaluate below correspond to a strict semiclassical expansion about some particular vacuum of the nonrelativistic BMN model.
%%%%%%%%%%%%%%%%%%%%%%%%%%%
\subsubsection{TSNC data}
We define the vielbeins corresponding to the NATD $ AdS $ (\ref{e2}) as
\begin{eqnarray}
\label{e10}
e^{\hat{0}}_{t}=L \cosh \varrho ~;~e^{\hat{1}}_{\varrho}=L~;~e^{\hat{2}}_{\eta}=\frac{2\alpha'}{L \sinh\varrho}~;~e^{\hat{3}}_{\chi}=\sqrt{\Delta}~;~e^{\hat{4}}_{\xi}=\sqrt{\Delta}\sin\chi ,
\end{eqnarray}
where we define, $ \Delta(\varrho , \eta)=\frac{4L^2 \alpha'^2 \eta^2 \sinh^2\varrho}{(16 \alpha'^2 \eta^2 +L^4 \sinh^4\varrho)}$.

Given (\ref{e10}), our next task would be to decode TSNC data following \cite{Bidussi:2021ujm}
\begin{eqnarray}
\label{e11}
E_{\mu}^A = \tau_{\mu}^{A}+\frac{1}{2c^2}\pi_{\mu}^B \epsilon_B^A.
\end{eqnarray}

We introduce following nonrelativistic (NR) scaling \cite{Gomis:2005pg}
\begin{eqnarray}
\label{e12}
L=c^2 \ell ~;~ t= \frac{\tilde{t}}{c}~;~  \varrho = \frac{\tilde{\varrho}}{c}~;~\alpha' \eta =c \alpha'_{NR}\tilde{\eta}~;~ \chi = \tilde{\chi}~;~\xi = \tilde{\xi}.
\end{eqnarray}

Uising (\ref{e12}), we figure out the following
\begin{eqnarray}
\label{e13}
e^{\hat{0}}=e^{\hat{0}}_t dt=cE^0_t dt=c \ell (1+\frac{\varrho ^2}{2 c^2})dt,
\end{eqnarray}
where for simplicity we remove tildes on the r.h.s. of (\ref{e13}). We follow this procedure for the rest of the analysis as well.

Comparing with (\ref{e11}), we find
\begin{eqnarray}
\tau_t^0 = \ell ~;~\pi_t^1 = -\ell \varrho^2.
\end{eqnarray}

Next, we note down
\begin{eqnarray}
e^{\hat{1}} = e^{\hat{1}}_{\varrho}d\varrho = c E_{\varrho}^1 d\varrho =c \ell d\varrho,
\end{eqnarray}
which clearly reveals
\begin{eqnarray}
\tau_{\varrho}^1 =\ell ~;~\pi_{\varrho}^0 =0.
\end{eqnarray}

The transverse vielbeins, on the other hand, read as
\begin{eqnarray}
e^{\hat{2}}&=&e^{\hat{2}}_{\eta}d\eta = e^{2}_{\eta}d\eta =\frac{2\alpha_{NR}'}{\ell \varrho} d\eta ,\\
e^{\hat{3}}&=& e^{\hat{3}}_{\chi}d \chi = e^{3}_{\chi} d\chi =\frac{2\alpha_{NR}' \eta}{\ell \varrho}d\chi ,\\
e^{\hat{4}}&=& e^{\hat{4}}_{\xi}d \xi = e^{4}_{\xi} d\xi =\frac{2\alpha_{NR}' \eta}{\ell \varrho}\sin\chi d\xi.
\end{eqnarray}

Finally, a direct computation of the non-longitudinal component \cite{Bidussi:2021ujm} of the Kalb-Ramond field yields
\begin{eqnarray}
B^{(NR)}_{\chi \xi}= \frac{1}{2}(e^{3}_{\chi}\pi_{\xi}^3 -e^{4}_{\xi}\pi_{\chi}^4)=-\frac{16 \alpha'^3_{NR}\eta^3 \sin\chi}{\ell^4 \varrho^4}+\mathcal{O}(c^{-2}),
\end{eqnarray}
where we define the NR scaling for NS fields as, $ B^{(NR)}_{\chi \xi} =c B_{\chi \xi} $.
%%%%%%%%%%%%%%%%%%%%%%%%%%%%%%%%%%
\subsubsection{NR sigma model}
The Ployakov action for the NR strings is given by 
\begin{eqnarray}
S^{(NR)}_{P}=-\frac{\ell^4}{4\pi \alpha'^3_{NR}}\int d^2\sigma \mathcal{L}^{(NR)}_{P},
\end{eqnarray} 
where the NR Lagrangian density could be formally expressed as  \cite{Bidussi:2021ujm}
\begin{eqnarray}
\label{e22}
\mathcal{L}^{(NR)}_{P} =\sqrt{- \gamma}\gamma^{\alpha \beta} e_{\mu}^a e_{\nu}^b \partial_{\alpha}X^{\mu}\partial_{\beta}X^{\nu}\delta_{ab}+\eta_{AB}(\tau_{\mu}^A\pi_{\nu}^B - \tau_{\nu}^A\pi_{\mu}^B)\dot{X}^{\mu}X'^{\nu}\nonumber\\
+\delta_{ab}(e_{\mu}^a \pi_{\nu}^b -e_{\nu}^a \pi_{\mu}^b)\dot{X}^{\mu}X'^{\nu}+\zeta \varepsilon^{\alpha \beta}e_{\alpha}^+ \tau_{\mu}^+ \partial_{\beta}X^{\mu}+\bar{\zeta}\varepsilon^{\alpha \beta}e_{\alpha}^- \tau_{\mu}^-\partial_{\beta}X^{\mu},
\end{eqnarray}
where we denote $ \dot{X}^{\mu}=\partial_0X^{\mu} $ and $ X'^{\mu}=\partial_1 X^{\mu} $ together with $\partial_{\alpha} = \frac{\partial}{\partial \sigma^{\alpha}} $ ($ \alpha =0,1 $).

Moreover, here $ \zeta $ and $ \bar{\zeta} $ are the world-sheet scalars \cite{Bergshoeff:2018yvt} together with the world-sheet vielbeins $ e_{\alpha}^{\pm}=e_{\alpha}^0 \pm e_{\alpha}^1 $ and $ \tau_{\mu}^{\pm}=\tau_{\mu}^0 \pm  \tau_{\mu}^1$. We also introduce the world-sheet metric
\begin{eqnarray}
\gamma_{\alpha \beta}=e_{\alpha}^{\tilde{a}} e_{\beta}^{\tilde{b}} \eta_{\tilde{a}\tilde{b}},
\end{eqnarray}
where $ \tilde{a}, \tilde{b} =0,1$ correspond to the world-sheet tangent space indices.

Using conformal gauge, one could further simplify (\ref{e22}) as
\begin{eqnarray}
\label{ee2.24}
S^{(NR)}_P =\frac{\sqrt{\lambda_{NR}}}{\pi}\int d^2 \sigma \mathcal{L}^{(NR)}_P ~;~ \sqrt{\lambda_{NR}}=\frac{\ell^2}{\alpha'_{NR}}.
\end{eqnarray}

The corresponding NR Lagrangian density reads as\footnote{We consider that the NR string is extended along the longitudinal direction $ \varrho = \varrho (\sigma^1) $ of the bulk target spacetime while it can have a non-trivial dynamics along the transverse directions of the TSNC manifold.}
\begin{eqnarray}
\label{e25}
\mathcal{L}^{(NR)}_P = \frac{1}{\varrho^2}(\dot{\eta}^2 -\eta'^2)+\frac{\eta^2}{\varrho^2}(\dot{\chi}^2 -\chi'^2)+\frac{\eta^2}{\varrho^2}\sin^2\chi (\dot{\xi}^2 -\xi'^2)-\frac{\ell^4}{4 \alpha'^2_{NR}}\dot{t}\varrho^2 \varrho' \nonumber\\
-\frac{8 \alpha'_{NR}}{\ell^2}\frac{\eta^3}{\varrho^4}\sin\chi (\dot{\xi}\chi' -\dot{\chi}\xi') +\frac{\ell^3}{4\alpha'^2_{NR}}(\zeta -\bar{\zeta})(\dot{t}-\varrho'),
\end{eqnarray}
where we fix the world-sheet gauge symmetries by choosing the matrix $ e_{\alpha}^{\tilde{a}} $ to be diagonal together with $ e_0 ^0 =1=e_1 ^1 $. 

The above Lagrangian density (\ref{e25}) represents the NS-NS sector of the NR type IIA strings propagating over NATD $ AdS $ background and therefore serves as the starting point for the rest of our analysis.
%%%%%%%%%%%%%%%%%%%%%%%%%%%%%%%%%%
\subsubsection{Nonrelativistic spinning string solution}
To explore semiclassical spectrum of these NR spinning strings, we propose the following embedding for the world-sheet degrees of freedom
\begin{eqnarray}
\label{e2.26}
t =\kappa \sigma^0 ~;~ \eta = \eta(\sigma^1)~;~\varrho = \varrho (\sigma^1)~;~\chi = \chi (\sigma^1)~;~\xi = \omega \sigma^0 ~;~\zeta =\zeta (\sigma^1)~;~\bar{\zeta}=\bar{\zeta}(\sigma^1).
\end{eqnarray}

The corresponding NR Lagrangian density turns out to be
\begin{eqnarray}
\label{EEE2.27}
\mathcal{L}^{(NR)}_P = -\frac{1}{\varrho^2}(\eta'^2 +\eta^2 \chi'^2 -\omega^2 \eta^2 \sin^2\chi)-\frac{\kappa \lambda_{NR}}{4}\varrho^2 \varrho' \nonumber\\
 - \frac{8 \omega}{\sqrt{\lambda_{NR}}}\frac{\eta^3}{\varrho^4}\sin\chi \chi' +\frac{\lambda_{NR}}{4\ell}(\zeta -\bar{\zeta})(\kappa -\varrho').
\end{eqnarray}

\paragraph{Equations of motion.} Below, we enumerate the resulting equations of motion
\begin{eqnarray}
\label{e28}
16 \omega \eta^3 \sin\chi \chi' +\lambda^{\frac{1}{2}}_{NR}\varrho^2 (\eta'^2 +\eta^2 \chi'^2 -\omega^2 \eta^2 \sin^2\chi)+\frac{\lambda^{\frac{3}{2}}_{NR}\varrho^5}{8 \ell}(\zeta' - \bar{\zeta}')&=&0,\\
\eta'' - 2\eta' \partial_{\sigma^1}\log\varrho -\eta \chi'^2 + \omega^2 \eta \sin^2\chi -\frac{12}{\lambda^{\frac{1}{2}}_{NR}}\omega \eta^2 \sin\chi \chi' &=&0,\\
\chi'' +2\chi' \partial_{\sigma^1}\log (\frac{\eta}{\varrho})+\omega^2 \sin\chi \cos\chi +\frac{4 \omega}{\lambda^{\frac{1}{2}}_{NR}}\frac{\eta}{\varrho^2}(3\partial_{\sigma^1}\log (\frac{\eta}{\varrho})-\partial_{\sigma^1}\log\varrho)\sin\chi &=&0,
\label{e30}
\end{eqnarray}
together with the solution for the radial coordinate as 
\begin{eqnarray}
\label{E2.31}
\varrho (\sigma^1)= \varrho_c +\mathcal{O}(\lambda^{-\frac{1}{2}}_{NR})= \kappa \sigma^1+\mathcal{O}(\lambda^{-\frac{1}{2}}_{NR}),
\end{eqnarray}
which results due to the equation motion of the world-sheet scalars $ \zeta $ or $ \bar{\zeta} $.

Considering the semiclassical limit $ \lambda_{NR}\gg 1 $, the simplest solutions that one could write down for the above set of equations (\ref{e28})-(\ref{e30}) are given by
\begin{eqnarray}
\zeta - \bar{\zeta} &=&\zeta_c +\mathcal{O}(\lambda_{NR}^{-1})~;~\chi =\chi_c =\frac{\pi}{2}+\mathcal{O}(\lambda^{-\frac{1}{4}}_{NR}),\\
\eta (\sigma^1)&=&\eta_c + \mathcal{O}(\lambda^{-\frac{1}{4}}_{NR})\nonumber\\
& =& -\sqrt{\frac{2}{\pi \omega^3 }}(\left(c_2 \sigma^1  \omega -c_1\right) \sin (\sigma^1  \omega )+\left(c_1 \sigma^1  \omega +c_2\right) \cos (\sigma^1  \omega ))+\mathcal{O}(\lambda^{-\frac{1}{4}}_{NR}).
\label{E2.33}
\end{eqnarray}

\paragraph{Conserved charges.} The (semi)classical energy ($ \mathcal{E}^{(cl)}_{NR} $) as well as the spin angular momentum ($ \mathcal{S}^{(cl)} $) associated with the NR sigma model turns out to be
\begin{eqnarray}
\label{e34}
\mathcal{E}^{(cl)}_{NR}& \sim &\frac{\lambda^{\frac{3}{2}}_{NR}}{2}\frac{\zeta_c}{\ell}(1-\frac{4\pi^2  \kappa^3}{3 \zeta_c }),\\
\mathcal{S}^{(cl)}_{(reg)}&\sim &\frac{ \lambda^{\frac{1}{2}}_{NR}}{\pi^3 }\frac{G (\omega)}{\omega^2 \kappa^2}-\mathcal{S}_{ct},
\label{e35}
\end{eqnarray}
together with
\begin{eqnarray}
G (\omega) = \left(c_1^2+c_2^2\right) \left(4 \pi ^2 \omega ^2-1\right)+\left(-2 \pi  c_2 c_1 \omega +c_1^2-c_2^2\right) \cos (4 \pi  \omega )\nonumber\\
+ \left(\pi  \left(c_1^2-c_2^2\right) \omega +2 c_1 c_2\right) \sin (4 \pi  \omega ).
\end{eqnarray}

Comparing (\ref{e34}) and (\ref{e35}) together we find the semiclassical spectrum of NR strings 
\begin{eqnarray}
\mathcal{E}^{(cl)}_{NR} \sim \mathcal{S}^{3(cl)}_{(reg)}.
\end{eqnarray}
%%%%%%%%%%%%%%%%%%%%%%%%%%%%%%%%%%%%
\subsubsection{Quantum corrections}
We now aim for going beyond the semiclassical approximation and in particular to estimate one loop correction to the NR stringy spectrum. To this end, we propose a general expansion of the following form
\begin{eqnarray}
\label{EE2.47}
X^a &=& X^a_c + \frac{X^a_f}{\lambda^{1/4}_{NR}}+ \cdots ,\\
X^A &=& X^A_c +\frac{X^A_f}{\lambda^{1/2}_{NR}}+ \cdots ,\\
\zeta - \bar{\zeta} &=&\zeta_c + \frac{\zeta_f}{\lambda_{NR}}+ \cdots,
\label{EE2.40}
\end{eqnarray}
where $ A=0,1 $ corresponds to longitudinal coordinates $ \lbrace t, \varrho \rbrace $. On the other hand, $ X^a $ could be any of the transverse world-sheet fields $ \lbrace \eta , \chi, \xi\rbrace $. Here, $\lbrace X^a_c , X^A_c \rbrace$ correspond to the semiclassical solutions those are listed above in (\ref{E2.31})-(\ref{E2.33}).

Using (\ref{EE2.47}), the NR action (\ref{ee2.24}) in fluctuations ($ \Psi_f $) turns out to be
\begin{eqnarray}
S^{(NR)}_f=\frac{1}{\pi}\int d^2\sigma \mathcal{L}^{(NR)}_f,
\end{eqnarray}
where the fluctuation Lagrangian density turns out to be
\begin{eqnarray}
\label{EE2.51}
\mathcal{L}^{(NR)}_f =-\frac{1}{\varrho^2_c}\left( \partial_{\alpha}\eta_f \partial^{\alpha}\eta_f - \frac{2\eta'^2_c \varrho_f}{\varrho_c}  \right) -\frac{\eta^2_c}{\varrho^2_c}(\partial_{\alpha}\chi_f \partial^{\alpha}\chi_f +\partial_{\alpha}\xi_f \partial^{\alpha}\xi_f)\nonumber\\
-\frac{\eta^2_c}{\varrho^2_c}\omega^2 \left( \frac{2 \varrho_f}{\varrho_c}-\frac{\eta^2_f}{\eta^2_c}\right)+\frac{4 \omega \eta_f \eta_c}{ \varrho^2_c}\dot{\xi}_f -\frac{\eta^2_c}{\varrho^2_c}\omega^2 \chi^2_f -\frac{\kappa}{4}\varrho'_f \varrho^2_f -\frac{\zeta_f \varrho'_f}{4 \ell},
\end{eqnarray}
where we choose to work with, $ t_f =0 $.

\paragraph{Equations of motion.} Equations of motion that readily follow from (\ref{EE2.51}) could be formally expressed as
\begin{eqnarray}
\label{EE2.52}
\square \eta_f - 2\eta'_f \partial_{\sigma^1}\log \varrho_c +\omega^2 \eta_f +2 \omega \eta_c\dot{\xi}_f& = &0,\\
\zeta'_f + \frac{8 \ell}{\varrho^3_c}(\eta'^2_c - \omega^2 \eta^2_c ) & =& 0,\\
\square \xi_f +2 \xi'_f \partial_{\sigma^1}\log \left( \frac{\eta_c}{\varrho_c}\right)-\frac{2\omega}{\eta_c}\dot{\eta}_f & =&0,\\
\square \chi_f +2 \chi'_f \partial_{\sigma^1}\log \left( \frac{\eta_c}{\varrho_c}\right)-\omega^2 \chi_f & =&0. 
\label{EE2.54}
\end{eqnarray}

It is easy to check that, the above set of equations (\ref{EE2.52})-(\ref{EE2.54}) are precisely the equations (\ref{e28})-(\ref{e30}) when expanded upto leading order in the quantum fluctuations. 

Considering a special case of fluctuations where
\begin{eqnarray}
\chi_f = \chi_f (\sigma^1)~;~\eta_f =\eta_f (\sigma^1),
\end{eqnarray}
it is straightforward to find the following solutions
\begin{eqnarray}
\eta_f &=& \frac{\kappa^2}{3}(\sigma^1)^3 + \mathcal{O}(\omega),\\
\chi_f & =&-\int \frac{\eta^2_c}{\varrho^2_c} d\sigma^1 + \mathcal{O}(\omega^2),
\end{eqnarray}
in the slow spinning limit namely, $ \omega \ll 1 $.

Finally, the equations of motion of $ \zeta_f $ and $ \varrho_f $ lead to the following set of solutions
\begin{eqnarray}
\varrho_f = \varrho_{fc}=\text{constant}~;~\zeta_f =-8 \ell \int \frac{\eta'^2_c}{\varrho^3_c}d\sigma^1 + \mathcal{O}(\omega^2).
\end{eqnarray}

\paragraph{Conserved charges.} The quantum one loop corrections to the energy and spin of the NR strings could be formally expressed as
\begin{eqnarray}
\mathcal{E}_{NR} &=& \mathcal{E}^{(cl)}_{NR} \left(1-\frac{2 \pi  \kappa^2}{\sqrt{\lambda}_{NR}}\frac{\varrho_{fc}}{\zeta_c}\frac{1}{(1-\frac{4\pi^2  \kappa^3}{3 \zeta_c })} +\cdots\right),\\
\mathcal{S}& =& \mathcal{S}^{(cl)}+\mathcal{O}(1/\lambda^{1/4}_{NR}).
\end{eqnarray}
%%%%%%%%%%%%%%%%%%%%%%%%%%%%%%%%%
\subsubsection{Liouvillian non-integrability}
\label{liouville}
The Liouvillian non-integrability criteria that we discuss below was first introduced by authors \cite{Basu:2011fw}-\cite{Stepanchuk:2012xi} in the context of relativistic strings which were further generalised for type IIA strings in \cite{Nunez:2018qcj} and very recently in the context of NR type IIB sigma models in \cite{Roychowdhury:2019olt}. 

The basic idea of Liouvillian non-integrability (which is based on the algorithm\footnote{See, the Appendix of \cite{Nunez:2018qcj} for a detailed discussion on the Kovacic algorithm.} due to Kovacic \cite{kova}) relies on the notion of classical integrability of Hamiltonian dynamics. The steps that one follows are essentially the following - (i) one has to choose an invariant plane \cite{Stepanchuk:2012xi} in the phase space, (ii) consider fluctuations those are normal to this plane and obtain the corresponding normal variational equation (NVE) and finally (iii) check whether the associated NVE allows solutions in terms of simple algebraic/harmonic functions, logarithms, exponentials etc. collectively known as Liouvillian form of solutions \cite{Basu:2011fw}.

In cases when the NVE admits a Liouvillian solution, the corresponding phase space configuration is termed as Liouville \emph{integrable}. In other words, the lack of such solutions essentially rules out the integrability of the associated phase space configuration. 

To summarise, unlike the standard Lax pair formulation of classical integrability, Kovacic algorithm and in particular the entire methodology of Liouvillian non-integrability can be used to disprove the classical integrability of a given phase space configuration. In other words, using the Kovacic formalism, one could therefore only conjecture about the integrability of a given string embedding. This is precisely the strategy that we follow here.

\paragraph{NVE.} To implement the algorithm in the present example of NR strings, we refer to the $ 1d $ reduction (\ref{e2.26}) and the resulting equations of motion in the strict semiclassical limit
\begin{eqnarray}
\label{e2.38}
\eta'' - 2\eta' \partial_{\sigma^1}\log\varrho -\eta \chi'^2 + \omega^2 \eta \sin^2\chi & \approx &0,\\
\chi'' +2\chi' \partial_{\sigma^1}\log (\frac{\eta}{\varrho})+\omega^2 \sin\chi \cos\chi & \approx &0.
\label{e2.39}
\end{eqnarray}

Notice that, $ \chi'' =\chi' =\chi =0 $ is an automatic solution of (\ref{e2.39}). Therefore, we choose $ \lbrace \chi =0 , p_{\chi}=0\rbrace $ to be the invariant plane in the phase space and consider fluctuations
\begin{eqnarray}
\chi \sim 0 + y (\sigma^1)~;~|y |\ll 1
\end{eqnarray}
those are normal to this plane.

Substituting the above choice into (\ref{e2.38}), we find
\begin{eqnarray}
\label{e2.41}
\eta (\sigma^1) \sim \frac{(\sigma^1) ^3}{3}.
\end{eqnarray}

Using (\ref{e2.41}), the corresponding NVE \cite{Nunez:2018qcj} finally turns out to be
\begin{eqnarray}
\label{e2.42}
y'' (\sigma^1)+\mathcal{P}(\sigma^1)y'(\sigma^1)+\mathcal{Q}(\sigma^1)y(\sigma^1) \approx 0,
\end{eqnarray}
where, the individual coefficients read as
\begin{eqnarray}
\mathcal{P}(\sigma^1) =\frac{4}{\sigma^1}~;~\mathcal{Q}(\sigma^1)=\omega^2.
\end{eqnarray}

The corresponding solution to the NVE (\ref{e2.42}) turns out to be
\begin{eqnarray}
\label{e2.44}
y(\sigma^1)=-\sqrt{\frac{2 }{\pi\omega ^3 (\sigma^1)^{6}}}(\left(c_2 \sigma^1  \omega -c_1\right) \sin (\sigma^1  \omega )+\left(c_1 \sigma^1  \omega +c_2\right) \cos (\sigma^1  \omega )).
\end{eqnarray}

The corresponding Schroedinger form \cite{Nunez:2018qcj} of the NVE turns out to be 
\begin{eqnarray}
\varpi' (\sigma^1)+\varpi^{2}(\sigma^1)=V(\sigma^1)=\frac{2}{(\sigma^1)^2}-\omega^2.
\end{eqnarray}

The corresponding solution
\begin{eqnarray}
\label{e2.46}
\varpi (\sigma^1)|_{\sigma^1 \ll 1}\sim -\frac{1}{\sigma^1 }+\sigma^{1}  \omega ^2,
\end{eqnarray}
behaves as a polynomial of degree 1 when expanded close to $ \sigma^1 \sim 0 $.

Combining both (\ref{e2.44}) and (\ref{e2.46}), we therefore conclude that NR strings those considered here are Liouville integrable.
%%%%%%%%%%%%%%%%%%%%%%%%%%%%%%%%%%%%%%%%%%%%%%%%%
\subsection{Nonrelativistic limit of ST background}
Following our previous approach, we now intend towards taking a NR limit \cite{Bidussi:2021ujm} of the ST background which is dual $ \mathcal{N}=2 $ linear quivers in $ 4d $ \cite{Lozano:2016kum}. To start with, we consider that the string is propagating over the $ 6d $ relativistic manifold of the following form
\begin{eqnarray}
ds^2 = -L^2 dt^2 +ds^2_{ST},
\end{eqnarray}
where we fix all the remaining coordinates of the $ 10d $ bulk spacetime.

The corresponding NS-NS two form is given by
\begin{eqnarray}
\label{E2.48}
B_2 = -\frac{16 \alpha'^3 \psi^3}{(16 \alpha'^2 \psi^2 +L^4 \cos^4\alpha)}\sin\theta d\theta \wedge d\varphi.
\end{eqnarray}
%%%%%%%%%%%%%%%%%%%%%%%%%%%%%%%%%%%%
\subsubsection{TSNC data}
We introduce relativistic vielbeins as
\begin{eqnarray}
\label{E2.49}
e^{\hat{0}}_{t}=L ~;~e^{\hat{1}}_{\alpha}=L~;~e^{\hat{6}}_{\beta}=L \sin\alpha ~;~e^{\hat{7}}_{\psi}=\frac{2\alpha'}{L \cos\alpha}~;~e^{\hat{8}}_{\theta}=\sqrt{\tilde{\Delta}}~;~e^{\hat{9}}_{\varphi}=\sqrt{\tilde{\Delta}}\sin\theta,
\end{eqnarray}
where we define, $ \tilde{\Delta}= \frac{4L^2 \alpha'^2 \psi^2 \cos^2\alpha}{(16 \alpha'^2 \psi^2 +L^4 \cos^4\alpha)}$.

We introduce the NR scaling \cite{Gomis:2005pg} of the following form
\begin{eqnarray}
\label{E2.50}
L=c^2 \ell ~;~ t= \frac{\tilde{t}}{c}~;~  \alpha = \frac{\tilde{\alpha}}{c}~;~\beta = \frac{\tilde{\beta}}{c}~;~\alpha' \psi =c^2 \alpha'_{NR}\tilde{\psi}~;~ \theta = \tilde{\theta}~;~\varphi = \tilde{\varphi}.
\end{eqnarray}

Using (\ref{E2.50}), we note down the following TSNC data corresponding to ST geometry
\begin{eqnarray}
\label{E2.51}
e^{\hat{0}}&=&c E^0_t dt ~;~ E^0_t = \ell = \tau_t^0 ,\\
e^{\hat{1}}&=&c E^1_{\alpha}d\alpha ~;~E^1_{\alpha}= \ell = \tau^1_{\alpha},\\
e^{\hat{6}}&=&e^{6}_{\beta}d\beta ~;~e^{6}_{\beta} =\ell \alpha ,\\
e^{\hat{7}}&=&e^{7}_{\psi}d\psi ~;~e^{7}_{\psi}=\frac{2\alpha'_{NR}}{\ell},\\
e^{\hat{8}}&=&e^{8}_{\theta}d\theta ~;~e^{8}_{\theta} = \frac{2\alpha'_{NR}\psi}{\ell},\\
e^{\hat{9}}&=&e^{9}_{\varphi}d\varphi ~;~e^{9}_{\varphi} = \frac{2\alpha'_{NR}\psi}{\ell}\sin\theta.
\end{eqnarray}

Finally, a NR limit of the NS-NS two form (\ref{E2.48}) reveals
\begin{eqnarray}
\label{E2.57}
B^{(NR)}_{\theta \varphi}= \frac{1}{2}(e^{8}_{\theta}\pi_{\varphi}^8 -e^{9}_{\varphi}\pi_{\theta}^9)=-16 \frac{\alpha'^3_{NR}}{\ell^4}\psi^3 \sin\theta +\mathcal{O}(c^{-4}),
\end{eqnarray}
where we define, $ B^{(NR)}_{\theta \varphi} =c^2  B_{\theta \varphi} $ under NR scaling.
%%%%%%%%%%%%%%%%%%%%%%%%%%%%%%%%%%
\subsubsection{NR sigma model}
With the above TSNC data (\ref{E2.51})-(\ref{E2.57}), we are now in a position to write down the corresponding sigma model action
\begin{eqnarray}
S^{(NR)}_P =\frac{\sqrt{\lambda_{NR}}}{\pi}\int d^2 \sigma \mathcal{L}^{(NR)}_P ,
\end{eqnarray}
in the NR limit of the ST background.

The associated Polyakov Lagrangian turns out to be\footnote{We consider that the NR string is extended along the longitudinal direction $ \alpha = \alpha (\sigma^1) $ of the bulk target spacetime while it can have a non-trivial dynamics along the transverse directions of the TSNC manifold.}
\begin{eqnarray}
\label{E2.59}
\mathcal{L}^{(NR)}_P=\lambda_{NR}\frac{\alpha^2}{4} (\dot{\beta}^2 -\beta'^2)+(\dot{\psi}^2 -\psi'^2)+\psi^2 (\dot{\theta}^2 -\theta'^2)+\psi^2 \sin^2\theta (\dot{\varphi}^2 -\varphi'^2)\nonumber\\
-\frac{8 \psi^3}{\sqrt{\lambda_{NR}}}\sin\theta (\dot{\varphi}\theta' -\dot{\theta}\varphi') +\frac{\lambda_{NR}}{4\ell} (\zeta - \bar{\zeta})(\dot{t}-\alpha').
\end{eqnarray}

Like in the previous example, the above Lagrangian density (\ref{E2.59}) represents F strings propagating over non-Lorentzian curve manifold that we identify as the TSNC limit of the ST background \cite{Lozano:2016kum}.
%%%%%%%%%%%%%%%%%%%%%%%%%%
\subsubsection{Nonrelativistic rotating string solutions}
To obtain the NR rotating string spectrum, we adopt the following F string embedding
\begin{eqnarray}
\label{E2.60}
t =\kappa \sigma^0 ~;~ \beta = \nu \sigma^0 ~;~\alpha = \alpha (\sigma^1)~;~\psi = \psi (\sigma^1),\nonumber\\
\varphi = \omega \sigma^0 ~;~ \theta =\theta (\sigma^1)~;~\zeta = \zeta (\sigma^1)~;~\bar{\zeta}=\bar{\zeta}(\sigma^1).
\end{eqnarray}

Using (\ref{E2.60}), the corresponding NR Lagrangian density (\ref{E2.59}) further simplifies as
\begin{eqnarray}
\label{E2.61}
\mathcal{L}^{(NR)}_P=\frac{\lambda_{NR}}{4}\nu^2 \alpha^2 - \psi'^2 - \psi^2 \theta'^2 +\omega^2 \psi^2 \sin^2\theta \nonumber\\
-\frac{8 \omega \psi^3}{\sqrt{\lambda_{NR}}}\sin\theta \theta' +\frac{\lambda_{NR}}{4\ell} (\zeta - \bar{\zeta})(\kappa -\alpha').
\end{eqnarray}

\paragraph{Equations of motion.} Below we enumerate the equations of motion that readily follow from (\ref{E2.61}).
\begin{eqnarray}
\psi'' -\psi (\theta'^2 - \omega^2 \sin^2\theta)-\frac{12 \psi^2 \omega}{\sqrt{\lambda_{NR}}}\sin\theta \theta'  =0,\\
\psi^2 (\theta'' +\omega^2 \sin\theta \cos\theta)+2\psi \psi' \theta' +\frac{12 \omega \psi^2 \psi'}{\sqrt{\lambda_{NR}}}\sin\theta =0,\\
2\nu^2 \alpha +\frac{1}{\ell}(\zeta' - \bar{\zeta}')=0,
\end{eqnarray}
together with the solutions
\begin{eqnarray}
\alpha = \alpha_c = \kappa \sigma^1,\\
\zeta - \bar{\zeta} = \zeta_c = -\nu^2 \kappa \ell (\sigma^1)^2.
\end{eqnarray}

Considering the strict semiclassical ($ \lambda_{NR}\gg 1 $) limit, the corresponding equations may be further simplified as
\begin{eqnarray}
\label{E2.65}
\psi'' -\psi (\theta'^2 - \omega^2 \sin^2\theta) \approx 0,\\
\psi^2 (\theta'' +\omega^2 \sin\theta \cos\theta)+2\psi \psi' \theta' \approx 0.
\label{E2.66}
\end{eqnarray}

Clearly, $ \theta = \theta_c = \frac{\pi}{2} +\mathcal{O}(\lambda_{NR}^{-\frac{1}{4}})$ solves the $ \theta $ equation (\ref{E2.66}). Substituting this into (\ref{E2.65}) we finally obtain
\begin{eqnarray}
\psi (\sigma^1 )\sim \psi_c \sim \sin (\omega \sigma^1),
\end{eqnarray}
where the picture that we have in mind is that of a NR closed string extended along the holographic axis ($ \psi $) \cite{Lozano:2016kum} and thereby also wrapping it.

\paragraph{Conserved charges.} Below, we enumerate all the conserved charges that readily follow from the NR sigma model (\ref{E2.59}). We first note down the energy of the NR string using the semiclassical approximation
\begin{eqnarray}
\mathcal{E}^{(cl)}_{NR} \sim \frac{2 \lambda^{3/2}_{NR}}{3}\pi^2 \nu^2 \kappa.
\end{eqnarray}

The angular momenta of rotation, on the other hand, turn out to be
\begin{eqnarray}
\mathcal{J}^{(cl)}_{\nu} &\sim & \frac{4 \lambda^{3/2}_{NR}}{3}\pi^2 \nu \kappa^2 ,\\
\mathcal{J}^{(cl)}_{\omega} &\sim & 2 \lambda^{1/2}_{NR}\omega.
\end{eqnarray}

Therefore, the most general form of the dispersion relation is formally expressed as
\begin{eqnarray}
\mathcal{E}^{(cl)}_{NR}\sim c_1 \mathcal{J}^{(cl)}_{\nu} +c_2 \mathcal{J}^{3(cl)}_{\omega},
\end{eqnarray}
where $ c_{1,2} $ being the coefficients of the expansion.
%%%%%%%%%%%%%%%%%%%%%%%%%%%%%%%%%%%
\subsubsection{Beyond semiclassical limit}
We estimate beyond semiclassical corrections to the rotating string spectrum following similar expansion as given in (\ref{EE2.47})-(\ref{EE2.40}). Here, $ \lbrace \beta, \psi , \theta , \varphi \rbrace $ are the transverse coordinates. On the other hand, $ \alpha $ acts as the longitudinal coordinate.

The corresponding Lagrangian density in fluctuations reads as
\begin{eqnarray}
\mathcal{L}^{(NR)}_f =-\frac{\alpha^2_f}{4}\partial_{\alpha}\beta_f \partial^{\alpha}\beta_f -\partial_{\alpha}\psi_f \partial^{\alpha}\psi_f -\psi^2_c \partial_{\alpha}\theta_f \partial^{\alpha}\theta_f -\psi^2_c \partial_{\alpha}\varphi_f \partial^{\alpha}\varphi_f \nonumber\\
-\psi^2_c \omega^2 \theta^2_f + 4\omega \psi_c \psi_f  \dot{\varphi}_f + \omega^2 \psi^2_f - \frac{\zeta_f \alpha'_f}{4 \ell}.
\end{eqnarray}

The resulting equations of motion are given by
\begin{eqnarray}
-\alpha_f\partial_{\alpha}\beta_f \partial^{\alpha}\beta_f +\frac{\zeta'_f}{2\ell}& =&0,\\
\square \beta_f + 2 \partial_{\alpha}\log \alpha_f \partial^{\alpha}\beta_f &=&0,\\
\square \psi_f + 2 \omega \psi_c \dot{\varphi}_f + \omega^2 \psi_f &=&0,\\
\square \theta_f + 2 \partial_{\sigma^1}\log\psi_c \theta'_f  - \omega^2 \psi_c \theta_f & =&0.
\end{eqnarray}

In the presence of these added fluctuations, the quantum corrections to the conserved charges are estimated as
\begin{eqnarray}
\mathcal{E}_{NR}& = & \mathcal{E}^{(cl)}_{NR}\left( 1+\frac{3}{8 \pi^3 \nu^2 \kappa \ell \lambda_{NR}}\int \zeta_f d\sigma^1+ \cdots \right), \\
\mathcal{J}_{\nu}&=&\mathcal{J}^{(cl)}_{\nu}\left(1+\frac{3}{8 \pi^3 \nu \lambda^{1/4}_{NR}}\int (\sigma^1)^2 \dot{\beta}_f d\sigma^1 + \cdots \right) ,\\
\mathcal{J}_{\omega}&=&\mathcal{J}^{(cl)}_{\omega}+\mathcal{O}(\lambda_{NR}^{-\frac{1}{4}}).
\end{eqnarray}
%%%%%%%%%%%%%%%%%%%%%%%%%%%%%%%%%%%
\subsubsection{Liouvillian non-integrability}
To check the Liouvillian (non)integrability, as before, one needs to boil down the equations of motion (\ref{E2.65})-(\ref{E2.66}) into the corresponding form of the NVE. 

To start with, we identify that $ \theta =\theta' = \theta'' =0 $ is a natural solution of (\ref{E2.66}). Which thereby naturally fixes the invariant plane in the phase space to be of the form $ \lbrace \theta =0 , p_{\theta}=0\rbrace $. The fluctuations about this plane would naturally correspond to the desired NVE.

Substituting the above choice into (\ref{E2.65}), we find
\begin{eqnarray}
\psi =a \sigma^1 +b.
\end{eqnarray}

Finally, considering a fluctuation of the form $ \delta \theta \sim y (\sigma^1) $, we arrive at the NVE of the form
\begin{eqnarray}
\label{E2.76}
y'' (\sigma^1)+\mathcal{P}(\sigma^1)y'(\sigma^1)+\mathcal{Q}(\sigma^1)y(\sigma^1) \approx 0,
\end{eqnarray}
with the coefficients of the form
\begin{eqnarray}
\mathcal{P}(\sigma^1) = \frac{2a}{a \sigma^1 +b}~;~\mathcal{Q}(\sigma^1)=\omega^2 .
\end{eqnarray}

The corresponding Liouvillian form of solution could be expressed as
\begin{eqnarray}
y (\sigma^1) = \textit{Re}\left[\frac{2 c_1 \omega  e^{-i \sigma^1  \omega }-i c_2 e^{i \sigma^1  \omega }}{2 a \sigma^1  \omega +2 b \omega } \right].
\end{eqnarray}

Finally, the Schroedinger form \cite{Nunez:2018qcj} of the NVE (\ref{E2.76}) takes the following expression
\begin{eqnarray}
\varpi' (\sigma^1)+\varpi^{2}(\sigma^1)=V(\sigma^1)=-\omega^2 ,
\end{eqnarray}
whose solution can be expressed as a polynomial of degree 1 
\begin{eqnarray}
\varpi (\sigma^1) = \omega ^2 \left(c_1-\sigma^1 \right) +\mathcal{O}(\omega^3),
\end{eqnarray}
in an expansion in the frequency ($ \omega \ll 1 $) - which thereby ensures the Liouville integrability of the sigma model.
%%%%%%%%%%%%%%%%%%%%%%%%%%%%%%%%%%%%%
\subsection{T-duality for NR sigma model}
Our discussion in this Section is based on the canonical formulation of T-duality transformation rules as originally developed by authors in \cite{Alvarez:1994wj}-\cite{Alvarez:1994dn}. Following their construction, we would eventually come up with a set of transformation rules (under T-duality) in the NR limit of strings propagating over NATD $ AdS_5 \times S^5 $.
%%%%%%%%%%%%%%%%%%%%%%%%%%%%%%%%%%%%%%%%%%
\subsubsection{T-duality for NR spinning strings}
The corresponding NR sigma model is described by the Lagrangian density (\ref{e25}) where the F-string is propagating over the TSNC limit of NATD background with $ \xi $ being the direction of isometry along which we apply T-duality transformation. 

To start with, we rewrite the NR Lagrangian density (\ref{e25}) as
\begin{eqnarray}
\mathcal{L}^{(NR)}_P =g_{\eta \eta}(\dot{\eta}^2 -\eta'^2)+g_{\chi \chi}(\dot{\chi}^2 -\chi'^2)+g_{\xi \xi} (\dot{\xi}^2 -\xi'^2)-\frac{\ell^4}{4 \alpha'^2_{NR}}\dot{t}\varrho^2 \varrho' \nonumber\\
-\frac{\ell^2}{2 \alpha'^2_{NR}}B^{(NR)}_{\xi \chi} (\dot{\xi}\chi' -\dot{\chi}\xi')+\frac{\ell^3}{4\alpha'^2_{NR}}(\zeta -\bar{\zeta})(\dot{t}-\varrho'),
\end{eqnarray}
which corresponds to spinning strings extended along the radial direction ($ \varrho $) of the target space.

The corresponding generating function \cite{Alvarez:1994wj}-\cite{Alvarez:1994dn} is given by
\begin{eqnarray}
G= \frac{1}{2}\int d\sigma^1 (\partial_{\sigma^1}\xi \tilde{\xi} - \xi \partial_{\sigma^1}\tilde{\xi}),
\end{eqnarray}
where $ \tilde{\xi} $ is the dual coordinate.

The canonically conjugate momenta ($ p_{\xi} $) and the dual momenta ($ \tilde{p}_{\xi} $) are expressed as
\begin{eqnarray}
\label{e2.49}
p_{\xi}= - \partial_{\sigma^1}\tilde{\xi}\equiv - \tilde{\xi}'~;~\tilde{p}_{\xi}= - \partial_{\sigma^1}\xi \equiv - \xi'.
\end{eqnarray}

Finally, we note down the canonically conjugate momenta associated with the NR sigma model
\begin{eqnarray}
\label{e2.52}
p_t &=&\frac{\ell^3}{4\alpha'^2_{NR}}(\zeta -\bar{\zeta}) - \frac{\ell^4}{4 \alpha'^2_{NR}}\varrho^2 \varrho',\\
p_{\eta}&=&2 g_{\eta \eta}\dot{\eta}~;~ p_{\chi} = 2 g_{\chi \chi}\dot{\chi}+\frac{\ell^2}{2 \alpha'^2_{NR}}B^{(NR)}_{\xi \chi}\xi', \\
p_{\xi} &=& 2 g_{\xi \xi}\dot{\xi}-\frac{\ell^2}{2 \alpha'^2_{NR}}B^{(NR)}_{\xi \chi} \chi' ,\\
p_{\varrho} &=& p_{\zeta}=p_{\bar{\zeta}}=0.
\label{e2.54}
\end{eqnarray}

The above relations (\ref{e2.52})-(\ref{e2.54}) can be inverted to obtain
\begin{eqnarray}
\label{e2.55}
\dot{\eta} &=&\frac{p_{\eta}}{2g_{\eta \eta}}~;~ \dot{\chi}=\frac{1}{2g_{\chi \chi}}(p_{\chi} -\frac{\ell^2}{2 \alpha'^2_{NR}}B^{(NR)}_{\xi \chi}\xi') ,\\
\dot{\xi}&=& \frac{1}{2 g_{\xi \xi}}(p_{\xi}+\frac{\ell^2}{2 \alpha'^2_{NR}}B^{(NR)}_{\xi \chi} \chi').
\label{e2.56}
\end{eqnarray}

Using (\ref{e2.55})-(\ref{e2.56}), the \emph{canonical} Hamiltonian density turns out to be
\begin{eqnarray}
\mathcal{H}=p_a \dot{X}^a +p_t \dot{t}+p_{\xi}\dot{\xi}-\mathcal{L}^{(NR)}_P,
\end{eqnarray}
which can be further simplified as
\begin{eqnarray}
\mathcal{H}=\frac{p^2_{\eta}}{4 g_{\eta \eta}}+\frac{p^2_{\chi}}{4g_{\chi \chi}}+\frac{p^2_{\xi}}{4g_{\xi \xi}}+g_{\eta \eta}\eta'^2 +g_{\chi \chi}\chi'^2 +g_{\xi \xi}\xi'^2 
+\varrho'(p_t +\frac{\ell^4}{4 \alpha'^2_{NR}}\varrho^2 \varrho')\nonumber\\
+\frac{\ell^4}{16 \alpha'^4_{NR}}\left( \frac{\chi'^2 B^{2(NR)}_{\xi \chi}}{g_{\xi \xi}}+\frac{\xi'^2 B^{2(NR)}_{\xi \chi}}{g_{\chi \chi}}\right) +\frac{\ell^2}{4 \alpha'^2_{NR}}\left( \frac{p_{\xi}}{g_{\xi \xi}}B^{(NR)}_{\xi \chi}\chi' -\frac{p_{\chi}}{g_{\chi \chi}}B^{(NR)}_{\xi \chi}\xi'\right) .
\end{eqnarray}

The total Hamiltonian of the system is defined as
\begin{eqnarray}
H=\int d\sigma^1 \mathcal{H}.
\end{eqnarray}

The dual Hamiltonian is obtained by replacing the original variables $(p_{\xi}, \xi)$ in terms of their dual ones (\ref{e2.49}). This leads to the following T dual Hamiltonian
\begin{eqnarray}
\tilde{\mathcal{H}}=\frac{p^2_{\eta}}{4 g_{\eta \eta}}+\frac{p^2_{\chi}}{4g_{\chi \chi}}+\frac{\tilde{\xi}'^2}{4g_{\xi \xi}}+g_{\eta \eta}\eta'^2 +g_{\chi \chi}\chi'^2 +g_{\xi \xi}\tilde{p}^2_{\xi}
+\varrho'(p_t +\frac{\ell^4}{4 \alpha'^2_{NR}}\varrho^2 \varrho')\nonumber\\
+\frac{\ell^4}{16 \alpha'^4_{NR}}\left(  \frac{\chi'^2 B^{2(NR)}_{\xi \chi}}{g_{\xi \xi}}+\frac{\tilde{p}_{\xi}^2 B^{2(NR)}_{\xi \chi}}{g_{\chi \chi}}\right) -\frac{\ell^2}{4 \alpha'^2_{NR}}\left( \frac{B^{(NR)}_{\xi \chi}}{g_{\xi \xi}}\chi' \tilde{\xi}' - \frac{p_{\chi}}{g_{\chi \chi}}B^{(NR)}_{\xi \chi}\tilde{p}_{\xi}\right) .
\end{eqnarray}

In order to decode the metric as well as the NS-NS fields corresponding to the T dual background, one has to get back to the Lagrangian formulation of the T dual string. This is achieved by means of the equations of motion using the T dual Hamiltonian
\begin{eqnarray}
\tilde{H}=\int d\sigma^1 \tilde{\mathcal{H}}.
\end{eqnarray}

A straightforward computation reveals
\begin{eqnarray}
\label{e2.67}
\dot{\tilde{\xi}}&=&\left( 2g_{\xi \xi}+\frac{\ell^4}{8 \alpha'^4_{NR}}\frac{B^{2(NR)}_{\xi \chi}}{g_{\chi \chi}}\right) \tilde{p}_{\xi}+\frac{\ell^2}{4\alpha'^2_{NR}}\frac{B^{(NR)}_{\xi \chi}}{g_{\chi \chi}}p_{\chi}~;~\dot{t} = \varrho' ,\\
\dot{\eta} &=&\frac{p_{\eta}}{2g_{\eta \eta}}~;~\dot{\chi}=\frac{p_{\chi}}{2g_{\chi \chi}}+\frac{\ell^2}{4 \alpha'^2_{NR}}\frac{B^{(NR)}_{\xi \chi}}{g_{\chi \chi}}\tilde{p}_{\xi}.
\label{e2.68}
\end{eqnarray}

On solving (\ref{e2.67}) and (\ref{e2.68}) we find
\begin{eqnarray}
\label{e2.69}
p_{\eta} &=&2g_{\eta \eta}\dot{\eta}~;~p_{\chi}=2g_{\chi \chi}\dot{\chi}-\frac{\ell^2}{2\alpha'^2_{NR}}B^{(NR)}_{\xi \chi}\tilde{p}_{\xi} ,\\
\tilde{p}_{\xi}&=&\frac{\dot{{\tilde{\xi}}}}{2g_{\xi \xi}}-\frac{\ell^2}{4\alpha'^2_{NR}}\frac{B^{(NR)}_{\xi \chi}}{•
g_{\xi \xi}}\dot{\chi}.
\label{e2.70}
\end{eqnarray}

Finally, we note down the T dual Lagrangian for the NR spinning string solution (\ref{EEE2.27})
\begin{eqnarray}
\label{e2.72}
\tilde{\mathcal{L}}_P &=& \tilde{g}_{\eta \eta}(\dot{\eta}^2 -\eta'^2 )+\tilde{g}_{\chi \chi}(\dot{\chi}^2 -\chi'^2) +\tilde{g}_{\tilde{\xi}\tilde{\xi}}(\dot{\tilde{\xi}}^2  -\tilde{\xi}'^2)-\tilde{g}_{\varrho \varrho}\varrho'^2 +\tilde{g}_{\chi \tilde{\xi}}(\dot{\chi}\dot{\tilde{\xi}}-\chi' \tilde{\xi}'),\nonumber\\
&\equiv & \sqrt{-\gamma}\gamma^{\alpha\beta}\tilde{g}_{\mu \nu}\partial_{\alpha}X^{\mu}\partial_{\beta}X^{\nu}.
\end{eqnarray}

Here the T dual components of the metric read as
\begin{eqnarray}
\label{e2.122}
\tilde{g}_{\eta \eta}&=&g_{\eta \eta}=\frac{1}{\varrho^2}~;~\tilde{g}_{\varrho \varrho}=\frac{\ell^4}{4 \alpha'^2_{NR}}\varrho^2 ,\\
\tilde{g}_{\chi \chi}&=& g_{\chi \chi}+\frac{\ell^4}{16 \alpha'^4_{NR}}\frac{B^{2(NR)}_{\chi \xi}}{g_{\xi \xi}},\\
\tilde{g}_{\chi \tilde{\xi}}&=&\frac{\ell^2}{4 \alpha'^2_{NR}}\frac{B^{(NR)}_{\chi \xi}}{g_{\xi \xi}}~;~\tilde{g}_{\tilde{\xi}\tilde{\xi}}=\frac{1}{4g_{\xi \xi}}.
\label{e2.124}
\end{eqnarray}

In the above $ \tilde{g}_{\mu \nu} $ represents the metric of the corresponding general relativity background with a trivial time component characterised by $ \tilde{g}_{tt}=\pm 1 $ along with $ t=\sigma^0 $.
%%%%%%%%%%%%%%%%%%%%%%%%%%%%%%%%%%%%%%%
\subsubsection{T-duality for NR rotating strings}
To find out T-duality rules for NR rotating strings (\ref{E2.60}), we adopt similar procedure as that of spinning strings. To start with, we write down the NR sigma model in the following form
\begin{eqnarray}
\mathcal{L}^{(NR)}_P=g_{\beta \beta}(\dot{\beta}^2-\beta'^2)+g_{\psi \psi}(\dot{\psi}^2 -\psi'^2)+g_{\theta \theta}(\dot{\theta}^2 -\theta'^2)+g_{\varphi \varphi}(\dot{\varphi}^2-\varphi'^2)\nonumber\\
-\frac{\sqrt{\lambda_{NR}}}{2\alpha'_{NR}}B^{(NR)}_{\varphi \theta}(\dot{\varphi}\theta' -\dot{\theta}\varphi')+ \frac{\lambda_{NR}}{4\ell} (\zeta - \bar{\zeta})(\dot{t}-\alpha'),
\end{eqnarray}
where $ \varphi $ is the direction of isometry along which we apply T-duality.

The corresponding generating function is given by
\begin{eqnarray}
G= \frac{1}{2}\int d\sigma^1 (\partial_{\sigma^1}\varphi \tilde{\varphi} - \varphi \partial_{\sigma^1}\tilde{\varphi}),
\end{eqnarray}
where $ \tilde{\varphi} $ is the dual coordinate.

Like before, this results in the dual momenta of the form
\begin{eqnarray}
\label{e2.127}
p_{\varphi}= - \partial_{\sigma^1}\tilde{\varphi}\equiv - \tilde{\varphi}'~;~\tilde{p}_{\varphi}= - \partial_{\sigma^1}\varphi \equiv - \varphi'.
\end{eqnarray}

Next, we note down the canonically conjugate momenta which read as
\begin{eqnarray}
\label{e2.128}
p_t &=&\frac{\lambda_{NR}}{4\ell} (\zeta - \bar{\zeta}),\\
p_{\beta}&=&2 g_{\beta \beta}\dot{\beta}~;~ p_{\psi} = 2g_{\psi \psi}\dot{\psi}, \\
p_{\theta} &=& 2g_{\theta \theta}\dot{\theta}+\frac{\sqrt{\lambda_{NR}}}{2\alpha'_{NR}}B^{(NR)}_{\varphi \theta}\varphi' ,\\
p_{\varphi} &=& 2g_{\varphi \varphi}\dot{\varphi}-\frac{\sqrt{\lambda_{NR}}}{2\alpha'_{NR}}B^{(NR)}_{\varphi \theta}\theta',\\
p_{\alpha}&=&0~;~p_{\zeta}=p_{\bar{\zeta}}=0.
\label{e2.131}
\end{eqnarray}

The above relations (\ref{e2.128})-(\ref{e2.131}) could be inverted to obtain the velocities
\begin{eqnarray}
\dot{\beta}&=&\frac{p_{\beta}}{2g_{\beta \beta}}~;~\dot{\psi}=\frac{p_{\psi}}{2g_{\psi \psi}},\\
\dot{\theta}&=&\frac{1}{2g_{\theta \theta}}(p_{\theta} -\frac{\sqrt{\lambda_{NR}}}{2\alpha'_{NR}}B^{(NR)}_{\varphi \theta}\varphi'),\\
\dot{\varphi}&=&\frac{1}{2g_{\varphi \varphi}}(p_{\varphi}+\frac{\sqrt{\lambda_{NR}}}{2\alpha'_{NR}}B^{(NR)}_{\varphi \theta}\theta').
\end{eqnarray}

The corresponding Hamiltonian density turns out to be
\begin{eqnarray}
\mathcal{H}=\frac{p^2_{\beta}}{4g_{\beta \beta}}+\frac{p^2_{\psi}}{4g_{\psi \psi}}+\frac{p^2_{\theta}}{4g_{\theta \theta}}+\frac{p^2_{\varphi }}{4g_{\varphi \varphi}}+\alpha' p_t + g_{\beta \beta}\beta'^2 +g_{\theta \theta}\theta'^2 
+g_{\psi \psi}\psi'^2 +g_{\varphi \varphi}\varphi'^2 \nonumber\\
+\frac{\lambda_{NR}}{16\alpha'^2_{NR}}\left( \frac{B^{2(NR)}_{\varphi \theta}}{g_{\varphi \varphi}}\theta'^2 + \frac{B^{2(NR)}_{\varphi \theta}}{g_{\theta \theta}}\varphi'^2 \right) +\frac{\sqrt{\lambda_{NR}}}{4 \alpha'_{NR}}\left( \frac{p_{\varphi}}{g_{\varphi \varphi}}B^{(NR)}_{\varphi \theta}\theta' -\frac{p_{\theta}}{g_{\theta \theta}}B^{(NR)}_{\varphi \theta}\varphi' \right).
\end{eqnarray}

Using (\ref{e2.127}), the corresponding dual Hamiltonian turns out to be
\begin{eqnarray}
\tilde{\mathcal{H}}=\frac{p^2_{\beta}}{4g_{\beta \beta}}+\frac{p^2_{\psi}}{4g_{\psi \psi}}+\frac{p^2_{\theta}}{4g_{\theta \theta}}+\frac{\tilde{\varphi}'^2}{4g_{\varphi \varphi}}+\alpha' p_t + g_{\beta \beta}\beta'^2 +g_{\theta \theta}\theta'^2 
+g_{\psi \psi}\psi'^2 +g_{\varphi \varphi}\tilde{p}_{\varphi}^2 \nonumber\\
+\frac{\lambda_{NR}}{16\alpha'^2_{NR}}\left( \frac{B^{2(NR)}_{\varphi \theta}}{g_{\varphi \varphi}}\theta'^2 + \frac{B^{2(NR)}_{\varphi \theta}}{g_{\theta \theta}}\tilde{p}_{\varphi}^2 \right) -\frac{\sqrt{\lambda_{NR}}}{4 \alpha'_{NR}}\left( \frac{\tilde{\varphi}'}{g_{\varphi \varphi}}B^{(NR)}_{\varphi \theta}\theta' -\frac{p_{\theta}}{g_{\theta \theta}}B^{(NR)}_{\varphi \theta}\tilde{p}_{\varphi}\right).
\end{eqnarray}

The corresponding equations of motion can be inverted to obtain
\begin{eqnarray}
\label{e2.136}
p_{a}&=&2g_{ab}\dot{X}^b~;~(a,b = \beta , \psi )~;~\dot{t}=\alpha',\\
\tilde{p}_{\varphi}&=&\frac{\dot{\tilde{\varphi}}}{2g_{\varphi \varphi}}-\frac{\sqrt{\lambda_{NR}}}{4 \alpha'_{NR}}\frac{B^{(NR)}_{\varphi \theta}}{g_{\varphi \varphi}}\dot{\theta},\\
p_{\theta}&=&2 g_{\theta \theta}\dot{\theta} - \frac{\sqrt{\lambda_{NR}}}{2 \alpha'_{NR}}B^{(NR)}_{\varphi \theta}\tilde{p}_{\varphi}.
\label{e2.137}
\end{eqnarray}

Using (\ref{e2.136})-(\ref{e2.137}), the dual Lagrangian density for NR rotating strings turns out to be
\begin{eqnarray}
\tilde{\mathcal{L}}_P=\tilde{g}_{ab}(\dot{X}^a \dot{X}^b - X'^a X'^b)+ \tilde{g}_{\theta \theta}(\dot{\theta}^2 -\theta'^2 ) +\tilde{g}_{\tilde{\varphi}\tilde{\varphi}}(\dot{\tilde{\varphi}}^2 -\tilde{\varphi}'^2)+\tilde{g}_{\tilde{\varphi}\theta}(\dot{\tilde{\varphi}}\dot{\theta} - \tilde{\varphi}'\theta'),
\end{eqnarray}
where we identify the T dual metric components of the general relativity background as
\begin{eqnarray}
\label{ee2.142}
\tilde{g}_{ab}&=&g_{ab}~;~a,b = \beta , \psi \\
\tilde{g}_{\theta \theta}&=& g_{\theta \theta}+\frac{\lambda_{NR}}{16\alpha'^2_{NR}}\frac{B^{2(NR)}_{\theta \varphi}}{g_{\varphi \varphi}},\\
\tilde{g}_{\tilde{\varphi}\theta}&=&\frac{\sqrt{\lambda_{NR}}}{4 \alpha'_{NR}}\frac{B^{(NR)}_{\theta \varphi }}{g_{\varphi \varphi}}~;~\tilde{g}_{\tilde{\varphi}\tilde{\varphi}} = \frac{1}{4g_{\varphi \varphi}}.
\label{e2.144}
\end{eqnarray}
%%%%%%%%%%%%%%%%%%%%%%%%%%%%%%%%%
\subsubsection{QFT observables}
One can now estimate various field theory entities using the metric components (\ref{ee2.142})-(\ref{e2.144}) of the general relativity spacetime and conjecture that they are the same for the NR counterpart as the later is connected with the former through a transverse T-duality.

For the purpose of our present analysis, we estimate the central charge \cite{Macpherson:2014eza} of the dual QFT using the metric components (\ref{ee2.142})-(\ref{e2.144}). To begin with, we write down the general relativity background as
\begin{eqnarray}
\label{ee2.145}
ds^2 =a(R, x^i)(dx^2_{1,3}+b(R,x^i )dR^2) + \tilde{g}_{ij}dx^i dx^j ,
\end{eqnarray}
where $ \lbrace i , j\rbrace $ are the coordinates of the internal manifold. Here, we understand that the internal manifold is associated with some $ AdS_5 $ factor of the full $ 10d $ general relativity spacetime. This stems from the fact that the $ AdS_5 $ factor of the parent NATD background remains untouched during the NR reduction as well as under T-duality transformations. 

Following \cite{Lozano:2016kum}, \cite{Macpherson:2014eza} we first compute the volume of the internal manifold which for the present example turns out to be
\begin{eqnarray}
\mathcal{V}_{int}=\int_0^{2\pi}d\beta \int_0^{\pi} d\theta \int_0^{2\pi}d\tilde{\varphi}\int_0^{\pi N_5}d\psi \sqrt{a^3\det \tilde{g}_{ij}},
\end{eqnarray}
where $ N_5 $ is the number of NS5 branes within an interval $ \psi =0 $ and $ \psi = \pi N_5 $ \cite{Lozano:2016kum}.

A straightforward computation reveals
\begin{eqnarray}
\sqrt{\det \tilde{g}_{ij}}=\frac{\sqrt{\lambda_{NR}}\alpha_0}{4 \sin\theta}\sqrt{1-\frac{48 \psi^2}{\lambda_{NR}}}.
\end{eqnarray}

Setting, $ a=\frac{4R^2}{L^2_{ADS}} $ and $ b=\frac{L^4_{ADS}}{R^4} $ we finally obtain
\begin{eqnarray}
\label{e2.148}
\mathcal{V}_{int}&=&\sqrt{\lambda_{NR}} \frac{2\alpha_0 R^3}{L^3_{ADS}}\int_0^{2\pi}d\beta \int_{\theta_{min}}^{\theta_{max}} \frac{d\theta}{\sin\theta} \int_0^{2\pi}d\tilde{\varphi}\int_0^{\pi N_5}d\psi \sqrt{1-\frac{48 \psi^2}{\lambda_{NR}}}\nonumber\\
&\simeq &\frac{16\sqrt{3}\pi^4 \alpha_0 R^3 N^2_5}{L^3_{ADS}}  \log \left( \frac{\tan\left( \frac{\theta_{max}}{2}\right) }{\tan\left( \frac{\theta_{min}}{2}\right)}\right)\equiv \frac{\alpha_0 \mathcal{V}_c R^3 N^2_5}{L^3_{ADS}}, 
\end{eqnarray}
where we fix one of the directions of the target space as $ \alpha =\alpha_0 $.

Using (\ref{e2.148}), the central charge finally turns out to be
\begin{eqnarray}
\label{centralnst}
c_{NRST}=\frac{9L^4_{ADS}}{G_{10}}\frac{b^{3/2} \mathcal{H}^{7/2}}{\mathcal{H}'^3}=\frac{\alpha_0 \mathcal{V}_c }{8 G_{10}}L_{ADS}^7 N^2_5 \rightarrow \infty,
\end{eqnarray}
where we denote $ \mathcal{H}= \mathcal{V}^{2}_{int}$ and $ G_{10}=8 \pi^5 g_s^2 \alpha'^4 $ \cite{Lozano:2016kum}.
%%%%%%%%%%%%%%%%%%%%%%%%%%%%%%%%%%%%%%%
\section{Adding flavour branes}
The ST background (\ref{e3}) that we have considered so far corresponds to dual $ \mathcal{N}=2 $ linear quivers in $ 4d $ those are infinitely extended. This is due to the fact that there are no flavour branes to complete these quivers. A systematic completion of $ \mathcal{N}=2 $ linear quivers has been achieved by adding flavour D6 branes into the bulk type IIA construction \cite{ReidEdwards:2010qs}-\cite{Lozano:2016kum}.

The formulation is based on using a different set of coordinates and thereby redefining the $ \mathcal{N}=2 $ background in terms of a potential function $ V(\sigma , \eta) $ \cite{ReidEdwards:2010qs}-\cite{Nunez:2018qcj}
\begin{eqnarray}
\label{e2.142}
ds^2_{IIA} &=& 4f_1 (\sigma , \eta)ds^2_{AdS_5} +f_2 (\eta , \sigma)(d\sigma^2 +d\eta^2)+f_3(\eta , \sigma)d\Omega_2 (\chi , \xi)+f_4 (\sigma , \eta)d\beta^2\\
B_2 &=&f_5 (\sigma , \eta)\sin\chi d\chi \wedge d\xi\\
ds^2_{AdS_5}&=&-dt^2 \cosh^2\rho + d\rho^2 + \sinh^2\rho (d\theta^2 +\cos^2\theta d\phi^2_1 +\sin^2\theta d\phi^2_2),\\
f_1 &=&\left(\frac{2\dot{V}-\ddot{V}}{\partial^2_{\eta}V} \right)^{1/2}~,~ f_2 = f_1\frac{2\partial^2_{\eta}V}{\dot{V}}~,~f_3 = f_1\frac{2\partial^2_{\eta}V \dot{V}}{\Delta}\\
f_4 &=&f_1 \frac{4\partial^2_{\eta}V \sigma^2}{2\dot{V}-\ddot{V}}~,~f_5 = 2\left(\frac{\dot{V}\partial_{\eta}\dot{V}}{\Delta}-\eta \right),~\Delta = (2\dot{V}-\ddot{V})\partial^2_{\eta}V +(\partial_{\eta}\dot{V})^2,
\end{eqnarray}
that satisfies Laplace equation of electrostatics with a given charge density\footnote{Here we denote, $ \dot{V}=\sigma\partial_{\sigma}V $ and $ \ddot{V}=\sigma^2 \partial^2_{\sigma}V +\sigma \partial_{\sigma}V $.} $ \lambda (\eta) $
\begin{eqnarray}
\partial_{\sigma}(\sigma \partial_{\sigma}V)+\sigma \partial^2_{\eta}V=0~;~\lambda (\eta)=\sigma \partial_{\sigma}V(\sigma, \eta)|_{\sigma =0}.
\label{e2.146}
\end{eqnarray}

The correct quantization for charges requires that the charge density $ \lambda (\eta) $ must satisfy the following boundary conditions: (i) it must vanish at $ \eta =0 $ and (ii) it must be a piece wise linear function namely, $ \lambda = a_i \eta +q_i $ with some slope $ a_i $ \cite{ReidEdwards:2010qs}-\cite{Aharony:2012tz}. 

The change in slope corresponds to the location of the flavour $ D6 $ branes along the field theory axis ($ \eta $). However, for supersymmetric quivers of finite length (for which one must satisfy $ \lambda (N_c)=0 $), the second boundary condition has beeen relaxed \cite{Maldacena:2000mw}.
%%%%%%%%%%%%%%%%%%%%%%%%%%
\subsection{ST revisited}
The choice of the Sfetsos-Thompson (ST) background \cite{Sfetsos:2010uq} corresponds to the choice of the potential function of the form \cite{Lozano:2016kum},
\begin{eqnarray}
V_{ST}(\sigma, \eta)=\eta \log\sigma - \eta \frac{\sigma^2}{2}+\frac{\eta^3}{3}~;~\lambda (\eta)=\eta
\label{e2.147}
\end{eqnarray}
which corresponds to a $  \mathcal{N} =2 $ linear quiver with linearly increasing rank.

We choose to work with the following background
\begin{eqnarray}
ds^2_{ST}&=&-4f_1 (\sigma , \eta)dt^2+f_2 (\eta , \sigma)(d\sigma^2 +d\eta^2)+f_3(\eta , \sigma)d\Omega_2 (\chi , \xi)+f_4 (\sigma , \eta)d\beta^2 ,\\
B_2 &=&f_5 (\sigma , \eta)\sin\chi d\chi \wedge d\xi.
\end{eqnarray}

In this new formulation, the old coordinates are related to the new set of coordinates as \cite{Lozano:2016kum} $ \psi = \frac{L^2}{\alpha'}\eta $ and $ \sigma = \sin\alpha $. In what follows, to carry out our analysis, we set $ \alpha =0 $ at the beginning which thereby corresponds to an expansion of the potential function near $ \sigma \sim 0 $ \cite{Nunez:2018qcj}. Finally, comparing with our previous calculations we identify $ \theta =\chi $ and $ \varphi = \xi $.

A straightforward computation reveals the metric components of the following form
\begin{eqnarray}
\label{3.10}
f_1 (\sigma \sim 0, \eta) &=&1 ~;~f_2 (\sigma \sim 0, \eta)= 4,\\
\label{3.11}
f_3 (\sigma \sim 0, \eta) &=&\frac{4 \eta ^2}{4 \eta ^2+1}~;~f_4 (\sigma \sim 0 , \eta)=0 ,\\
\label{3.12}
f_5 (\sigma \sim 0, \eta) &=&-\frac{8 \eta ^3}{4 \eta ^2+1}.
\end{eqnarray}

\paragraph{TSNC data.} The NR limit is defined using the following set of scaling
\begin{eqnarray}
\label{e2.150}
t = c \tilde{t} ~;~\eta = c \tilde{\eta}~;~ \chi = \tilde{\chi}~;~\xi = \tilde{\xi}~;~B^{(NR)}_2=\frac{B_2}{c},
\end{eqnarray}
where, as before we do not use tildes in the subsequent analysis.

Clearly, here $ t $ and $ \eta $ act as the longitudinal direction of the TSNC spacetime while the other two directions ($ \chi , \xi \in S^2 $) act as the transverse directions. In what follows, we consider that the NR string is extended along the holographic/ field theory direction $ \eta = \eta (\sigma^1) $ while it can have non-trivial dynamics along the transverse directions ($ \chi , \xi \in S^2 $). 

 Below we summarise the resulting TSNC data those follow from the NR scaling (\ref{e2.150})
\begin{eqnarray}
\tau_{t}^0 &=& 2 =\tau^1_{\eta},~;~e^{2}_{\chi}= 1~;~e^{3}_{\xi}=\sin\chi ,\\
 B^{(NR)}_{\chi \xi} &=&-2 \eta \sin\chi  =\frac{1}{2}(e^{2}_{\chi}\pi_{\xi}^2 -e^{3}_{\xi}\pi_{\chi}^3).
\end{eqnarray}

This yields the NR sigma model Lagrangian of the following form
\begin{eqnarray}
\label{e3.16}
\mathcal{L}_P^{(NR)}=g_{\chi \chi}(\dot{\chi}^2 -\chi'^2)+g_{\xi \xi}(\dot{\xi}^2 -\xi'^2)-2B^{(NR)}_{\xi  \chi}(\dot{\xi}\chi' - \dot{\chi}\xi')+2(\zeta -\bar{\zeta})(\dot{t}-\eta'),
\end{eqnarray}
where the metric of the NR target space is identified as, $ g_{\chi \chi}=1 $ and $ g_{\xi \xi}=\sin^2\chi $.

A direct comparison with (\ref{E2.59}) reveals that one could identify the dynamics (\ref{e3.16}) as quite analogous to that of (\ref{E2.59}) (in natural units) where $ \eta $ (previously $ \psi $) now playing the role of the longitudinal direction (previously $ \alpha $) along which the NR string is extended. A direct comparison further reveals the trivial identification of the transverse directions namely, $ \chi =\theta $ and $ \xi = \varphi $ which remain unaffected under NR scaling.

\paragraph{T-duality and QFT observables.} We now proceed towards discussing a general canonical algorithm for deriving T-duality rules using this new set of coordinates. The analysis is quite analogous as before which therefore allows us to quote only major steps here.

Like before we consider applying T-duality along the transverse isometry direction $ \xi $ of the target space manifold. This leads to the dual Hamiltonian density of the form
\begin{eqnarray}
\label{EE3.17}
\tilde{\mathcal{H}}=\frac{p^2_{\chi}}{4g_{\chi \chi}}+\frac{\tilde{\xi}'^2}{4g_{\xi \xi}}+\left( g_{\chi \chi}+\frac{B^{2(NR)}_{\xi  \chi}}{g_{\xi \xi}}\right) \chi'^2 +\left( g_{\xi \xi}+\frac{B^{2(NR)}_{\xi  \chi}}{g_{\chi \chi}}\right) \tilde{p}_{\xi}^2\nonumber\\
-\frac{B^{(NR)}_{\xi  \chi}\tilde{\xi}' \chi'}{g_{\xi \xi}}+\frac{B^{(NR)}_{\xi  \chi}p_{\chi}\tilde{p}_{\xi}}{g_{\chi \chi}}+p_t \eta'.
\end{eqnarray}

Using (\ref{EE3.17}), we note down the following conjugate momenta
\begin{eqnarray}
p_{\chi}&=&2g_{\chi \chi}\dot{\chi}-2B^{(NR)}_{\xi  \chi}\tilde{p}_{\xi},\\
\tilde{p}_{\xi}&=&\frac{1}{2g_{\xi \xi}}(\dot{\tilde{\xi}}-2 B^{(NR)}_{\xi  \chi} \dot{\chi}),\\
\dot{t}&=& \eta'.
\end{eqnarray}

Using these data, the dual Lagrangian finally turns out to be
\begin{eqnarray}
\tilde{\mathcal{L}}_P =\tilde{g}_{\chi \chi}(\dot{\chi}^2 -\chi'^2)+\tilde{g}_{\tilde{\xi} \tilde{\xi}}(\dot{\tilde{\xi}}^2 -\tilde{\xi}'^2)+g_{\tilde{\xi}\chi}(\dot{\chi}\dot{\tilde{\xi}}-\chi' \tilde{\xi}'),
\end{eqnarray}
where the individual metric components read as
\begin{eqnarray}
\label{st3.22}
\tilde{g}_{\chi \chi} &=& g_{\chi \chi}+\frac{B^{2(NR)}_{ \chi \xi }}{g_{\xi \xi}} ,\\
\tilde{g}_{\tilde{\xi} \tilde{\xi}} &=&\frac{1}{4 g_{\xi \xi}}~;~g_{\tilde{\xi}\chi} = \frac{B^{(NR)}_{\chi \xi}}{g_{\xi \xi}},
\label{st3.23}
\end{eqnarray}
which are precisely the previously obtained T-duality rules (\ref{ee2.142})-(\ref{e2.144}) while the remaining coordinates of the internal manifold being fixed.

It is now straightforward to compute the central charge using the above data (\ref{st3.22})-(\ref{st3.23}). A detailed computation reveals the volume of the internal manifold as
\begin{eqnarray}
\mathcal{V}_{int}&=&\frac{4 R^3 \mathcal{V}_2}{L^3_{ADS}} \int_{\chi_{min}}^{\chi_{max}} \frac{d\chi}{\sin \chi} \int_0^{2\pi} d\tilde{\xi}\int_{0}^{N_5} d\eta \sqrt{| 1 -12 \eta^2 |}\nonumber\\
& \simeq & \frac{8 \pi \sqrt{3} R^3 \mathcal{V}_2 N^2_5}{L^3_{ADS}}  \log \left( \frac{\tan\left( \frac{\chi_{max}}{2}\right) }{\tan\left( \frac{\chi_{min}}{2}\right)}\right),
\end{eqnarray}
where $ \mathcal{V}_2 $ is the two dimensional volume due to other two coordinates ($ \sigma , \beta $). 

Following similar steps as before (\ref{centralnst}) we obtain the central charge as
\begin{eqnarray}
c_{NRST}=\frac{9 L^4_{ADS}}{G_{10}}\frac{b^{3/2} \mathcal{H}^{7/2}}{\mathcal{H}'^3}=c_0 N^2_5 \rightarrow \infty.
\end{eqnarray}
where $ c_0 $ is an overall constant that can be easily read out from the above data.
%%%%%%%%%%%%%%%%%%%%%%%%%%%%%
\subsection{Single kinks}
\label{sk}
\begin{figure}
\includegraphics[scale=.74]{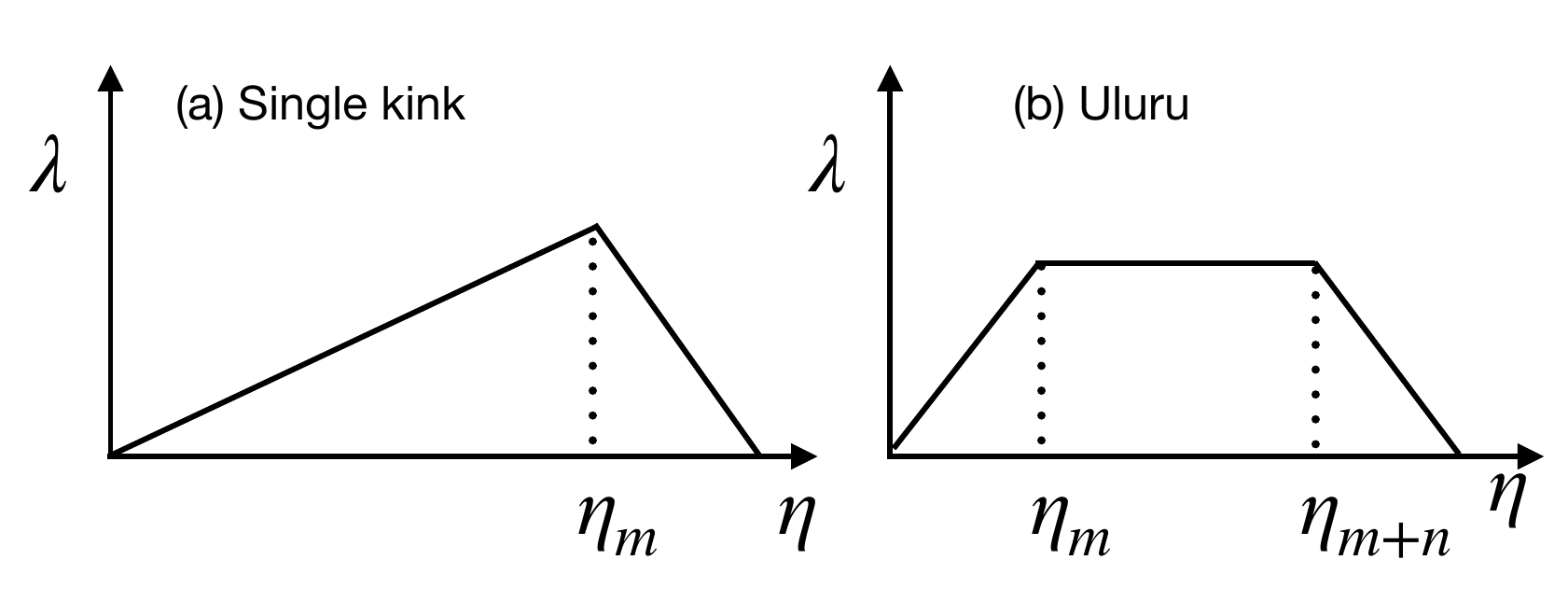}
  \caption{We plot the charge density ($ \lambda $) against the holographic direction ($ \eta $) which in the dual string theory description is identified as one of the longitudinal coordinates associated with the Lorentzian manifold.} \label{uluru}
\end{figure}

Single kink profile corresponds to $ \mathcal{N}=2 $ linear quivers with linearly increasing rank those are closed due to the addition of a $ SU((m+1)N_f) $ flavour group. In the dual string theory picture, this corresponds to placing flavour D6 branes precisely at $ \eta = \eta_m $ (see Fig. \ref{uluru}).  

The linearly increasing rank of the color group associated with the $ \mathcal{N}=2 $ quiver is characterised by a linear increase in the charge density $ \lambda (\eta) $ in the dual stringy picture. On the other hand, placing flavour nodes\footnote{Flavour nodes are placed in such a way so that the gauge anomaly cancellation is ensured \cite{Lozano:2016kum}.} correspond to a change in slope as shown in Fig. \ref{uluru}.

Single kink solutions \cite{Lozano:2016kum} are characterised by the following charge density
\begin{eqnarray}
\dot{V}_k(\eta , \sigma)=\frac{N_6}{2}\sum_{n=-\infty}^{\infty}(P+1)[\sqrt{\sigma^2 +(\eta +P -2n(1+P))^2}-\sqrt{(\eta -2 n (P+1)-P)^2+\sigma ^2}]\nonumber\\
+P[\sqrt{(\eta -2 n (P+1)-P-1)^2+\sigma ^2}-\sqrt{(\eta -2 n (P+1)+P+1)^2+\sigma ^2}],\nonumber\\
\label{e3.17}
\end{eqnarray}
where one considers placing $ N_6 $ flavor D6 branes at a location $ \eta = \eta_m =P $ of the type IIA manifold \cite{Nunez:2018qcj}.  

 Considering an expansion in the regime of small $ \sigma $ leads to the potential function of the following form
\begin{eqnarray}
V_k (\sigma \sim 0, \eta)= \eta N_6 \log\sigma +\frac{\eta N_6 \sigma^2}{4}\Lambda_{k}(\eta , P)
-\frac{\eta N_6 \sigma^2}{4}\frac{(P+1)}{(P^2 -\eta^2)},
\end{eqnarray}
where we express the function as
\begin{eqnarray}
\Lambda_{k}(\eta , P) =(P+1)\sum_{m=1}^{k}\frac{1}{(2m+(2m-1)P)^2-\eta^2}-\frac{1}{(2m+(2m+1)P)^2-\eta^2}\nonumber\\
+\frac{P}{(2k+1)^2(1 +P)^2 -\eta^2},
\end{eqnarray}
together with the fact that $ -k\leq n \leq  k$ where $ k $ is set to be as large as possible.

Like before, we compute the metric functions ($ f_i(\sigma , \eta) $) near $ \sigma \sim 0 $ which read
\begin{eqnarray}
\label{e3.20}
f_1(\sigma \sim 0, \eta)&= & 4f^{-1}_2(\sigma \sim 0, \eta),\\
f_2(\sigma \sim 0, \eta)&= &\frac{\sqrt{2} }{\sqrt{\frac{\eta  \left(P^2-\eta ^2\right)^3}{\sigma ^2 \left(\left(P^2-\eta ^2\right)^3 \left(\eta  \partial^2_{\eta}\Lambda_k (\eta , P )+2 \partial_{\eta}\Lambda_k (\eta , P )\right)-2 \eta  (P+1) \left(\eta ^2+3 P^2\right)\right)}}},\\
f_3(\sigma \sim 0, \eta)&= &  \eta^2 f_2(\sigma \sim 0, \eta),\\
f_4(\sigma \sim 0, \eta)&= &\sigma^2 f_2(\sigma \sim 0, \eta).
\label{e3.23}
\end{eqnarray}

%%%%%%%%%%%%%%%%%%%%%%%%%%%%%%
\subsubsection{TSNC data and the NR sigma model}
\label{flavb1}
Next, we carefully explore the metric functions (\ref{e3.20})-(\ref{e3.23}) under the NR scaling which we define below. This allows us to extract the corresponding TSNC data and thereby obtain a NR limit of the single kink profile as depicted above in (\ref{e3.17}).

A careful analysis in the large $ c $ limit of both the functions $ \partial_{\eta}\Lambda_k (\eta , P )|_{\eta = c \tilde{\eta}} $ and $ \partial^2_{\eta}\Lambda_k (\eta , P )|_{\eta = c \tilde{\eta}} $ leads to the following conclusion
\begin{eqnarray}
\label{3.33}
f_1(\sigma \sim \sigma_k , \tilde{\eta}) &=& \frac{2 \tilde{\eta}^2}{\sigma_k} ~;~ f_2(\sigma \sim \sigma_k , \tilde{\eta}) = \frac{2 \sigma_k}{\tilde{\eta}^2},\\
f_3(\sigma \sim \sigma_k , \tilde{\eta}) &=& 2c^2 \sigma_k ~;~f_4(\sigma \sim \sigma_k , \tilde{\eta}) = \frac{2c^4\sigma^3_k}{\tilde{\eta}^2},
\label{3.34}
\end{eqnarray}
where we scale $ \sigma \rightarrow \sigma_k= \frac{\sigma}{c^2}  $ while taking the large $ c $ limit with the ratio finite.

We introduce the following NR scaling for the rest of the coordinates
\begin{eqnarray}
t =c \tilde{t}~;~\chi = \frac{\tilde{\chi}}{c}~;~\xi =\tilde{\xi}~;~\beta = \frac{\tilde{\beta}}{c^2},
\end{eqnarray}
which leads to the following TSNC data
\begin{eqnarray}
\tau_t^0 &=& \frac{2\sqrt{2}\eta}{\sqrt{\sigma_k}}~;~\tau^1_{\eta}=\frac{\sqrt{2 \sigma_k}}{\eta},\\
e^2_{\chi}&=& \sqrt{2 \sigma_k}~;~e^3_{\xi}=\sqrt{2 \sigma_k} \chi ,\\
e^4_{\beta}&=&\frac{\sqrt{2}\sigma^{3/2}_k}{\eta},
\end{eqnarray}
where we remove tildes for simplicity.

To extract other TSNC data, one has to take a proper NR scaling of the transverse NS-NS flux. A careful analysis reveals the metric function of the following form
\begin{eqnarray}
f_5(\sigma \sim 0 , \eta) =\eta ^2 \sigma ^2 \left(-\eta  \partial^2_{\eta}\Lambda_k (\eta )-3 \partial_{\eta}\Lambda_k (\eta )+\frac{8 \eta  (P+1) P^2}{\left(P^2-\eta ^2\right)^3}\right)+\mathcal{O}(\sigma^3).
\end{eqnarray}

Using the NR scaling this finally leads to
\begin{eqnarray}
B^{(NR)}_{\chi \xi} =\frac{1}{2}(e^{2}_{\chi}\pi_{\xi}^2 -e^{3}_{\xi}\pi_{\chi}^3)=-\frac{8 \sigma^2_k  \chi}{\eta^3}P^2 (P+1) .
\end{eqnarray}

Using the above TSNC data, finally we are in a position to write down the NR Lagrangian for strings those are extended along the holographic axis $ \eta (\sigma^1) $
\begin{eqnarray}
\label{ee3.41}
\mathcal{L}^{(NR)}_P=g_{\chi \chi}(\dot{\chi}^2 -\chi'^2)+g_{\xi \xi}(\dot{\xi}^2 -\xi'^2)+g_{\beta \beta}(\dot{\beta}^2-\beta'^2)\nonumber\\
-2B^{(NR)}_{\xi  \chi}(\dot{\xi}\chi' - \dot{\chi}\xi')+(\zeta -\bar{\zeta})(\tau^0_t \dot{t}-\tau^1_{\eta}\eta'),
\end{eqnarray}
where the individual metric components are identified as
\begin{eqnarray}
\label{3.42}
g_{\chi \chi}=2 \sigma_k ~;~g_{\xi \xi}=2 \sigma_k \chi^2 ~;~ g_{\beta \beta}=\frac{2 \sigma^3_k }{\eta^2}.
\end{eqnarray}

Contrary to the previous ST example, here we notice a metric singularity at $ \eta \sim 0 $ which is due to the presence of NS5 branes in the parent NATD theory. The singularity at $ \eta \sim 0 $ is an artefact of the NR scaling $ \frac{\eta}{c} $ of the old coordinates in the large $ c $ limit. 

As a result of this scaling, some of the NS5 branes of the parent NATD theory are now located at $ \eta =0 $ causing a singularity there. Therefore, one has to put a lower cut-off $ \eta =\eta_{min} $ to avoid such singularities near the origin.

On the other hand, the upper bound $\eta  = \eta_{max} $ along the holographic axis is also scaled as $ \sim \frac{\eta_{max}}{c}\sim \frac{N_5}{c} $ where $ N_5 $ is the number of NS5 branes associated with the parent NATD background. These are precisely the new locations of the flavour D6 branes in the TSNC limit of the NATD backgrounds.

Considering a rotating string embedding of the form
\begin{eqnarray}
\chi = \chi (\sigma^1)~;~\beta = \text{const.} ~;~ \xi = \omega \sigma^0 ~;~ t = \sigma^0
\end{eqnarray}
the resulting equation of motion turns out to be
\begin{eqnarray}
\label{e3.35}
\chi'' + \omega^2 \chi -12 \omega \sigma_k P^2 (P+1)\frac{\eta'}{\eta^4}\chi =0.
\end{eqnarray}

The Lagrange multiplier fields $ \zeta $ and $ \bar{\zeta} $ on the other hand yields the following constraint
\begin{eqnarray}
\tau^1_{\eta}\eta' - \tau^0_t  = 0 ~\Rightarrow \eta (\sigma^1)=-\frac{\sigma_k}{\sigma_k c_1+2 \sigma^1 }.
\end{eqnarray}

\paragraph{Integrability.} To check Liouvillian (non)integrability we obtain the corresponding NVE by expanding (\ref{e3.35}) about $ \chi \sim 0 $, $ p_{\chi}\sim 0 $. This yields the following \emph{non-Liouvillian} solution
\begin{eqnarray}
y(\sigma^1)=c_3 D_{-\frac{\sigma_k \omega ^{3/2}}{8 \sqrt{6} P \sqrt{P+1}}-\frac{1}{2}}\left(\frac{\sqrt{P} \sqrt[4]{P+1} \sqrt[4]{\omega } \left(2 \sigma +\sigma_k c_1\right) \text{Root}\left[\text{$\#$1}^4-24\&,4\right]}{\sqrt{\sigma_k}}\right)\nonumber\\
+c_2 D_{\frac{\sigma_k\omega ^{3/2}}{8 \sqrt{6} P \sqrt{P+1}}-\frac{1}{2}}\left(\frac{2^{3/4} \sqrt[4]{3} \sqrt{P} \sqrt[4]{P+1} \sqrt[4]{\omega } \left(2 \sigma +\sigma_k c_1\right)}{\sqrt{\sigma_k}}\right)
\end{eqnarray}
in terms of special functions which we identify as ``parabolic cylindrical functions''. 

Notice that, these non-Liouvillian solutions are actually sourced due the $ B^{(NR)}_2  \sim P^2 (P+1)$ field in the process of NR reduction. A careful look further reveals that the NS sector eventually carries the information about the \emph{flavour} branes of the general relativity background which are therefore responsible for loosing integrability in the NR counterpart.  

This is quite reminiscent of what has been observed previously in the relativistic formulation of the theory \cite{Nunez:2018qcj} where one looses integrability due to the presence of the flavour branes in the bulk. To conclude, the Liouville \emph{non-integrablity} (in the sense of Kovacic  \cite{kova}) persists while considering the NR reduction in the presence of flavour D6 branes.
%%%%%%%%%%%%%%%%%%%%%%%%%%%%
\subsubsection{T-duality and QFT observables}
\label{3.2.2}
To find out T-duality rules, we start from the Lagrangian (\ref{ee3.41}) where the T-duality is applied along the transverse isometry direction $ \xi $. This leads to the dual Hamiltonian density of the form
\begin{eqnarray}
\label{3.47}
\tilde{\mathcal{H}}=\frac{p^2_{\chi}}{4g_{\chi \chi}}+\frac{p^2_{\beta}}{4g_{\beta \beta}}+\frac{\tilde{\xi}'^2}{4g_{\xi \xi}}+\left( g_{\chi \chi}+\frac{B^{2(NR)}_{\xi  \chi}}{g_{\xi \xi}}\right) \chi'^2 +\left( g_{\xi \xi}+\frac{B^{2(NR)}_{\xi  \chi}}{g_{\chi \chi}}\right) \tilde{p}_{\xi}^2\nonumber\\
+g_{\beta \beta}\beta'^2 -\frac{B^{(NR)}_{\xi  \chi}\tilde{\xi}' \chi'}{g_{\xi \xi}}+\frac{B^{(NR)}_{\xi  \chi}p_{\chi}\tilde{p}_{\xi}}{g_{\chi \chi}}+ \frac{p_t \tau_{\eta}^1}{\tau_t^0} \eta'.
\end{eqnarray}

Using (\ref{3.47}), the conjugate momenta are calculated as
\begin{eqnarray}
\label{3.48}
p_{\chi}&=&2g_{\chi \chi}\dot{\chi}-2B^{(NR)}_{\xi  \chi}\tilde{p}_{\xi},\\
\tilde{p}_{\xi}&=&\frac{1}{2g_{\xi \xi}}(\dot{\tilde{\xi}}-2 B^{(NR)}_{\xi  \chi} \dot{\chi}),\\
p_{\beta}&=& 2 g_{\beta \beta}\dot{\beta},\\
\dot{t}&=&\frac{\tau_{\eta}^1}{\tau_t^0} \eta'.
\label{3.51}
\end{eqnarray}

Using the above data (\ref{3.48})-(\ref{3.51}), the dual Lagrangian density finally turns out to be
\begin{eqnarray}
\tilde{\mathcal{L}}_P =\tilde{g}_{\beta \beta}(\dot{\beta}^2 -\beta'^2)+\tilde{g}_{\chi \chi}(\dot{\chi}^2 -\chi'^2)+\tilde{g}_{\tilde{\xi} \tilde{\xi}}(\dot{\tilde{\xi}}^2 -\tilde{\xi}'^2)+g_{\tilde{\xi}\chi}(\dot{\chi}\dot{\tilde{\xi}}-\chi' \tilde{\xi}'),
\end{eqnarray}
where the individual metric components read as
\begin{eqnarray}
\label{sk3.53}
\tilde{g}_{\chi \chi} &=& g_{\chi \chi}+\frac{B^{2(NR)}_{ \chi \xi }}{g_{\xi \xi}}~;~ \tilde{g}_{\beta \beta}=g_{\beta \beta} ,\\
\tilde{g}_{\tilde{\xi} \tilde{\xi}} &=&\frac{1}{4 g_{\xi \xi}}~;~g_{\tilde{\xi}\chi} = \frac{B^{(NR)}_{\chi \xi}}{g_{\xi \xi}}.
\label{sk3.54}
\end{eqnarray}

\paragraph{Central charge.} To compute the central charge, like before, one first computes the volume of the internal manifold
\begin{eqnarray}
\mathcal{V}_{int}&=&\frac{8 R^3 \sigma^{3/2}_k}{\sqrt{2}L^3_{ADS}} \int_{\chi_{min}}^{\chi_{max}} \frac{d\chi}{ \chi}\int_0^{2\pi} d\beta \int_0^{2\pi} d\tilde{\xi}\int_{\eta_{min}}^{\eta_{max}} \frac{d\eta}{\eta} \sqrt{ \frac{48 \sigma^2_k N^2_5 (N_5 -1)^4}{\eta^6}-1} \nonumber\\
& \simeq & \frac{128 \pi^2 R^3 \sigma^{5/2}_k N^3_5}{\sqrt{6}L^3_{ADS}\eta^3_{min}} \left( 1-\frac{\eta^3_{min}}{\eta^3_{max}}\right) \log \left( \frac{\chi_{max}}{\chi_{min}}\right)+\mathcal{O}( \sigma_k /N_5),
\end{eqnarray}
where we identify $ N_5 = P+1 $ together with, $ \eta_{max}\sim \frac{N_5}{c} $.

Finally, the central charge of the configuration is noted down to be
\begin{eqnarray}
\label{csk}
c_{NRSK}=\frac{9L^4_{ADS}}{G_{10}}\frac{b^{3/2} \mathcal{H}^{7/2}}{\mathcal{H}'^3}=\bar{c}_0 N^3_5,
\end{eqnarray}
where we define, $ \bar{c}_0 = \frac{16 \pi^2}{3\sqrt{6}}\frac{L^8_{ADS}}{G_{10}}\frac{\sigma^{5/2}_k}{\eta^3_{min}} \left( 1-\frac{\eta^3_{min}}{\eta^3_{max}}\right) \log \left( \frac{\chi_{max}}{\chi_{min}}\right)$.
%%%%%%%%%%%%%%%%%%%%%%%%%%
\subsection{Uluru profile}
\label{ul}
Uluru profile corresponds to associating flavour D6 branes at two distinct points ($ \eta_m $ and $ \eta_{m+n} $) along the holographic axis ($ \eta $) in the type IIA description (see Fig. \ref{uluru}). In the dual $ \mathcal{N}=2 $ SCFTs, this corresponds to adding flavour nodes in the long array of color nodes of the quiver and thereby satisfying the criteria of gauge anomaly cancellations \cite{ReidEdwards:2010qs}-\cite{Lozano:2016kum}.

The profile corresponds to a charge density of the form \cite{Lozano:2016kum}
\begin{eqnarray}
\label{e3.38}
\dot{V}_{u}(\sigma, \eta)=\frac{N_6}{2}\sum_{m=-\infty}^{\infty}\sum_{l=1}^{3}\sqrt{\sigma^2 +(\nu_l +2m(2P+K)-\eta)^2}\nonumber\\
-\sqrt{\sigma^2 +(\nu_l -2m(2P+K)+\eta)^2},
\end{eqnarray}
where we define $ \nu_1 = P $, $ \nu_2 = P+K $ and $ \nu_3=-(2P+K) $. The above background represents $ N_6 $ flavor D6 branes located at $ \eta =P $ and $ \eta =P+K $ along the holographic ($ \eta $) axis of the internal manifold.

Upon expanding the charge density (\ref{e3.38}) near $ \sigma \sim 0 $ we obtain the associated potential function
\begin{eqnarray}
V_{u}(\sigma \sim 0, \eta)= -\eta N_6 \log\sigma +\frac{\eta N_6 \sigma^2}{4(P^2 -\eta^2)}+\frac{\eta N_6 \sigma^2}{4}\Lambda_{u}(\eta , K, P),
\end{eqnarray}
where we denote the function as
\begin{eqnarray}
\Lambda_{u}(\eta , K, P)=\sum_{n=1}^{u}(-1)^{n+1}\left(\frac{1}{(n K+(2n-1)P)^2-\eta^2} -\frac{1}{(n K+(2n+1)P)^2-\eta^2}\right),
\end{eqnarray}
together with the bound $ -u \leq m \leq u $.

The resulting metric functions read as
\begin{eqnarray}
f_1(\sigma \sim 0, \eta)&= &4 f^{-1}_2(\sigma \sim 0, \eta),\\
f_2(\sigma \sim 0, \eta)&= & \frac{\sqrt{2} }{ \sqrt{\frac{\eta  \left(\eta ^2-P^2\right)^3}{\sigma ^2 \left(2 \eta ^3+\left(P^2-\eta ^2\right)^3 \left(\eta  \partial^2_{\eta}\Lambda_u (\eta , K, P)+2 \partial_{\eta}\Lambda_u (\eta , K, P)\right)+6 \eta  P^2\right)}}},\\
f_3(\sigma \sim 0, \eta)&= &\eta^2 f_2(\sigma \sim 0, \eta),\\
f_4(\sigma \sim 0, \eta)&= &\sigma^2 f_2(\sigma \sim 0, \eta).
\end{eqnarray}
%%%%%%%%%%%%%%%%%%%%%%%%%%%%%%
\subsubsection{TSNC data and the NR sigma model}
\label{flavb2}
The strategy that we follow now is quite similar in spirit as that of the single kink profile. After repeating those steps here we find
\begin{eqnarray}
\label{e3.45}
f_1(\sigma \sim \sigma_k , \tilde{\eta}) &=& \frac{2 \tilde{\eta}^2}{\sigma_k} ~;~ f_2(\sigma \sim \sigma_k , \tilde{\eta}) = \frac{2 \sigma_k}{\tilde{\eta}^2},\\
f_3(\sigma \sim \sigma_k , \tilde{\eta}) &=& 2c^2 \sigma_k ~;~f_4(\sigma \sim \sigma_k , \tilde{\eta}) = \frac{2c^4\sigma^3_k}{\tilde{\eta}^2},
\label{e3.46}
\end{eqnarray}
where as before we define $ \sigma_k =\frac{\sigma}{c^2}= \text{fixed} $ and $ \eta = c \tilde{\eta} $.

Clearly these metric coefficients (\ref{e3.45})-(\ref{e3.46}) are exactly identical to those obtained for the single kink profile. A straightforward computation, on the other hand reveals
\begin{eqnarray}
\label{e3.47}
f_5(\sigma \sim 0 , \eta) =\eta ^2 \sigma ^2 \left(3 \partial_{\eta}\Lambda_u +\eta  \left(\partial^2_{\eta}\Lambda_u +\frac{8 P^2}{\left(P^2-\eta ^2\right)^3}\right)\right)+\mathcal{O}(\sigma^3).
\end{eqnarray}

Using (\ref{e3.47}), the NR NS-NS two form could be formally expressed as
\begin{eqnarray}
B^{(NR)}_{\chi \xi} =\frac{1}{2}(e^{2}_{\chi}\pi_{\xi}^2 -e^{3}_{\xi}\pi_{\chi}^3)=-\frac{8 \sigma^2_k  \chi}{\eta^3}P^2  .
\end{eqnarray}

Qualitatively, we therefore have the identical scenario as that of the single kink profile which allows us to conclude that the Uluru profile is also non-integrable in the same sense as that of the single kink profile.
%%%%%%%%%%%%%%%%%%%%%%%%%%%%
\subsubsection{T-duality and QFT observables}
The T-duality rules for Uluru profile are the same as that of the single kink profile (\ref{sk3.53})-(\ref{sk3.54}) for obvious reasons- except for the fact that we now have a different $ B_2 $ field.

Taking all these into account, the volume of the internal manifold turns out to be
\begin{eqnarray}
\label{3.68}
\mathcal{V}_{int}&=&\frac{8 R^3 \sigma^{3/2}_k}{\sqrt{2}L^3_{ADS}} \int_{\chi_{min}}^{\chi_{max}} \frac{d\chi}{ \chi}\int_0^{2\pi} d\beta \int_0^{2\pi} d\tilde{\xi}\int_{\eta_{min}}^{\eta_{max}} \frac{d\eta}{\eta} \sqrt{   \frac{3 \sigma^2_k  (N_5 -K)^4}{\eta^6}-1} \nonumber\\
& \simeq & \frac{32 \pi^2 R^3 \sigma^{5/2}_k N^2_5}{\sqrt{6}L^3_{ADS}\eta^3_{min}}\left( 1-\frac{\eta^3_{min}}{\eta^3_{max}}\right)  \log \left( \frac{\chi_{max}}{\chi_{min}}\right)+\mathcal{O}(K \sigma_k /N_5),
\end{eqnarray}
where we identify $ N_5 = 2P+K $ together with, $ \eta_{max}\sim \frac{N_5}{c} $.

Using (\ref{3.68}), the central charge finally turns out to be
\begin{eqnarray}
\label{culu}
c_{NRUL}=\hat{c}_0 N^2_5,
\end{eqnarray}
where we denote, $ \hat{c}_0 = \frac{4 \pi^2}{3\sqrt{6}}\frac{L^8_{ADS}}{G_{10}}\frac{\sigma^{5/2}_k}{\eta^3_{min}} \left( 1-\frac{\eta^3_{min}}{\eta^3_{max}}\right)  \log \left( \frac{\chi_{max}}{\chi_{min}}\right)$.
%%%%%%%%%%%%%%%%%%%%%%%%%%%%%%%%%%%%%%%
\section{Remarks on generic Gaiotto-Maldacena backgrounds}
\label{GM}
We now generalise all the above results in terms of the single electrostatic potential function $ V(\sigma , \eta) $ that defines the generic Gaiotto-Maldacena (GM) class of geometries (\ref{e2.142})-(\ref{e2.146}). We consider a general expansion for the potential function close to $ \sigma \sim 0 $ \cite{Nunez:2018qcj}
\begin{eqnarray}
V(\sigma , \eta)=\mathcal{F}(\eta)+a \eta \log \sigma + \sum_{m=1}^{\infty}h_m (\eta)\sigma^{2m},
\end{eqnarray}
where we define
\begin{eqnarray}
4h_1 (\eta)=-\mathcal{F}''(\eta)~;~h_m (\eta)=-\frac{1}{4m^2}h''_{m-1}(\eta)~;~m=2,3,\cdots .
\end{eqnarray}
%%%%%%%%%%%%%%%%%%%%%%%%%%%%
\subsection{Large $ c $ limit}
Let us now consider the NR scaling as introduced previously
\begin{eqnarray}
\label{4.3}
\sigma_k =\frac{\sigma}{c^2}= \text{fixed} ~;~\eta = c \tilde{\eta}.
\end{eqnarray}

With (\ref{4.3}), we compute the following entities
\begin{eqnarray}
\label{4.4}
\dot{V}(\sigma \sim \sigma_k , \tilde{\eta})&=&   \tilde{\eta} a c -\frac{c^2}{2}\sigma^2_k \mathcal{F}^{(2)}( \tilde{\eta})-\sum_{m=2}^{\infty}\frac{c^{4m -2}}{2m}h^{(2)}_{m-1}( \tilde{\eta})\sigma_k^{2m},\\
\ddot{V}(\sigma \sim \sigma_k ,  \tilde{\eta})&=&-c^2\sigma^2_k \mathcal{F}^{(2)}( \tilde{\eta})-\sum_{m=2}^{\infty}c^{4m -2}h^{(2)}_{m-1}( \tilde{\eta})\sigma_k^{2m},\\
V''(\sigma \sim \sigma_k ,  \tilde{\eta})&=&-\frac{\mathcal{F}^{(4)}( \tilde{\eta})}{4}\sigma^2_k +\frac{\mathcal{F}^{(2)}( \tilde{\eta})}{c^2}-\sum_{m=2}^{\infty}c^{4(m-1)}\frac{h^{(4)}_{m-1}( \tilde{\eta})}{4m^2}\sigma^{2m}_k ,\\
\dot{V}'(\sigma \sim \sigma_k ,  \tilde{\eta})&=&a -\frac{c}{2}\mathcal{F}^{(3)}( \tilde{\eta})\sigma^2_k -\sum_{m=2}^{\infty}c^{4m-3}\frac{h^{(3)}_{m-1}( \tilde{\eta})}{2m }\sigma^{2m}_k ,
\label{4.7}
\end{eqnarray}
where prime corresponds to derivative w.r.t. $  \tilde{\eta} $. On the other hand, the superscripts count the number of $  \tilde{\eta} $ derivatives.

\paragraph{NR charge density.} In the strict limit of $ \sigma_k \rightarrow 0 $, the charge density \cite{Lozano:2016kum} in the TSNC limit of the GM background turns out to be
\begin{eqnarray}
\lambda_{NR}=\frac{\lambda}{c}= a  \tilde{\eta} .
\end{eqnarray}

Clearly, for ST, the $  \tilde{\eta} $ direction is unbounded which corresponds to a quantum mechanical model with infinite rank of the gauge group and infinite central charge. On the other hand, with flavour D6 branes, the holographic $  \tilde{\eta} $ direction for the GM background is bounded above which results in the $ \mathcal{N}=2 $ quivers with finite rank of the gauge group. 

While taking the TSNC limit, the location(s) of these flavour branes are shifted and rescaled as $ \sim \frac{ \eta_m}{c}\sim \frac{N_5}{c} \sim  \tilde{\eta}_{max} $. In other words, these flavour branes are now shifted more towards the origin of the holographic $  \tilde{\eta} $ axis. 

On a similar note, the location of the NS5 branes are also shifted as compared to their parent NATD theory and some of these branes are now located at the origin of the rescaled holographic axis. The presence of these NS5 branes are precisely reflected as the metric singularities of the form (\ref{3.42}). Due to the presence of these flavour branes in TSNC limit of GM backgrounds, one therefore expects to find a finite central charge for the dual QFT as computed above in (\ref{csk}) and (\ref{culu}).

Finally, we compute the metric elements and take their respective large $ c $ limits
\begin{eqnarray}
\label{4.9}
f_1(\sigma \sim \sigma_k ,  \tilde{\eta})&=& \sqrt{\frac{2 a c^3 \tilde{\eta}+\sum_{m=1}^{\infty}\frac{m}{m+1}c^{4(m+1)}h^{(2)}_m(\tilde{\eta})\sigma_k^{2(m+1)}}{\mathcal{F}^{(2)}(\tilde{\eta})-\frac{\mathcal{F}^{(4)}(\tilde{\eta})}{4}c^2 \sigma^2_k - \sum_{m=1}^{\infty}\frac{1}{4(m+1)^2}c^{4m+2}h^{(4)}_m(\tilde{\eta})\sigma_k^{2(m+1)}}},\\
f_2(\sigma \sim \sigma_k ,  \tilde{\eta})&=& 2 f_1\left( \frac{\mathcal{F}^{(2)}(\tilde{\eta})-\frac{\mathcal{F}^{(4)}(\tilde{\eta})}{4}c^2 \sigma^2_k - \sum_{m=1}^{\infty}\frac{1}{4(m+1)^2}c^{4m+2}h^{(4)}_m(\tilde{\eta})\sigma_k^{2(m+1)}}{\tilde{\eta} a c^3 -\frac{c^4}{2}\sigma^2_k \mathcal{F}^{(2)}( \tilde{\eta})-\sum_{m=1}^{\infty}\frac{c^{4(m+1) }}{2(m+1)}h^{(2)}_{m}( \tilde{\eta})\sigma_k^{2(m+1)}}\right),\\
 f_3(\sigma \sim \sigma_k ,  \tilde{\eta})&=&\frac{2f_1}{\hat{\Delta}}\left( \frac{\tilde{\eta}a}{c}\mathcal{F}^{(2)}(\tilde{\eta})-\frac{\tilde{\eta}a c}{4}\mathcal{F}^{(4)}(\tilde{\eta})\sigma^2_k - \frac{\sigma^2_k}{2}\mathcal{F}^{2(2)}+\mathcal{O}(\sigma^4_k c^2)\right),\\
 f_4(\sigma \sim \sigma_k ,  \tilde{\eta})&=& \frac{4 c^4 \sigma^2_k}{f_1},
\end{eqnarray}
where we denote
\begin{eqnarray}
\hat{\Delta}=\frac{2 a}{c}\tilde{\eta}\mathcal{F}^{(2)}(\tilde{\eta})-\frac{a c}{2}\tilde{\eta}\sigma^2_k \mathcal{F}^{(4)}(\tilde{\eta})-2 a c \tilde{\eta}\sum_{m=1}^{\infty}\frac{c^{4m}}{4(m+1)^2}h^{(4)}_{m}(\tilde{\eta})\sigma_k^{2(m+1)}\nonumber\\
+\frac{1}{c^4}\left(\mathcal{F}^{(2)}(\tilde{\eta})-\frac{c^2 \sigma^2_k}{4}\mathcal{F}^{(4)}(\tilde{\eta}) \right) \sum_{m=1}^{\infty}\frac{m c^{4(m+1)}}{(m+1)}h^{(2)}_{m}(\tilde{\eta})\sigma_k^{2(m+1)}+\mathcal{O}(\sigma_{k,m\geq 1}^{4(m+1)}).
\end{eqnarray}

Finally, we note down the NR limit of the NS-NS flux
\begin{eqnarray}
 f_5(\sigma \sim \sigma_k ,  \tilde{\eta})=\frac{2}{\hat{\Delta}}\left(\tilde{\eta}a^2 c -\frac{\tilde{\eta}ac^2}{2}\mathcal{F}^{(3)}(\tilde{\eta})\sigma^2_k -\tilde{\eta}a \sum_{m=1}^{\infty}\frac{c^{4m+2}}{2(m+1)}h^{(3)}_m(\tilde{\eta})\sigma^{2(m+1)}_k\right) \nonumber\\
 -\frac{c^2 \sigma^2_k}{\hat{\Delta}}\mathcal{F}^{(2)}(\tilde{\eta})\left(a-\frac{c}{2}\mathcal{F}^{(3)}( \tilde{\eta})\sigma^2_k -\sum_{m=1}^{\infty}c^{4m+1}\frac{h^{(3)}_{m}( \tilde{\eta})}{2(m+1) }\sigma^{2(m+1)}_k \right) -2 c \tilde{\eta}\nonumber\\
 -\sum_{m=1}^{\infty}\frac{c^{4m +2}}{2(m+1)}h^{(2)}_{m}( \tilde{\eta})\sigma_k^{2(m+1)}\left( a -\frac{c}{2}\mathcal{F}^{(3)}( \tilde{\eta})\sigma^2_k\right) +\mathcal{O}(\sigma^{4(m+1)}_{k,m\geq 1}).
\end{eqnarray}
%%%%%%%%%%%%%%%%%%%%%%%%%%%%
\subsection{Some examples}
As a quick check, for ST background (\ref{e2.147}), we identify
\begin{eqnarray}
\mathcal{F}(\tilde{\eta})&=&\frac{c^3 \tilde{\eta}^3}{3}~;~a =1~;~\Rightarrow\mathcal{F}^{(2)}(\tilde{\eta})=2c^3 \tilde{\eta},\\
h_1(\tilde{\eta})&=&-\frac{c \tilde{\eta}}{2}~;~h_{m\geq 2}(\tilde{\eta})=0, 
\end{eqnarray}
which upon substitution into (\ref{4.9}), yields $ f_1( \sigma_k \sim 0 ,  \tilde{\eta}) =1$ as found in (\ref{3.10}). Following a similar line of arguments, one can verify $ f_2( \sigma_k \sim 0 ,  \tilde{\eta}) =4$.

Using similar steps, one finds the remaining metric components as
\begin{eqnarray}
f_3 |_{\sigma_k =0}&=&2f_1 \frac{ \frac{\tilde{\eta}}{c}\mathcal{F}^{(2)}(\tilde{\eta})}{\frac{2 }{c}\tilde{\eta}\mathcal{F}^{(2)}(\tilde{\eta})}=1~;~f_4 |_{\sigma_k =0}=0,\\
f_5 |_{\sigma_k =0}&=&\frac{2 \tilde{\eta}c}{\frac{2 \tilde{\eta}}{c}\mathcal{F}^{(2)}(\tilde{\eta})}-2 c \tilde{\eta} \approx -2c \tilde{\eta}+\mathcal{O}(c^{-2}).
\end{eqnarray}
These are precisely the results obtained previously in (\ref{3.11}) and (\ref{3.12}) along with the substitution $ \eta = c \tilde{\eta} $, in the large $ c $ limit.

On a similar note, for single kink profile, one can identify the function
\begin{eqnarray}
h_1 (\eta)=\frac{\eta  }{4}\Lambda_{k}(\eta , P) -\frac{\eta }{4}\frac{(P+1)}{(P^2 -\eta^2)}~;~a=1,\\
\mathcal{F}^{(2)}(\eta) \sim -h_1(\eta)~;~\mathcal{F}^{(4)}(\eta)\sim -h_1''(\eta),
\end{eqnarray}
which together with the scaling (\ref{4.3}) reproduces the metric functions (\ref{3.33})-(\ref{3.34}) in the limit, $ \sigma_k \sim 0 $.
%%%%%%%%%%%%%%%%%%%%%%%%%%%%%%%%%%%%%%%%%%%%%%%%%%
\section{Concluding remarks and the limits of $ \mathcal{N}=2 $ SCFTs}
\label{conclud}
Before we proceed further, let us summarise the key results of the paper. The present work deals with a detailed analysis on TSNC limits \cite{Bidussi:2021ujm} of Gaiotto-Maldacena class of geometries \cite{Gaiotto:2009gz} of type IIA supergravity those preserve $ \mathcal{N}=2 $ supersymmetry\footnote{As our analysis reveals, in the large $ c $ limit of the ST background, the associated cloak function ($ \tau_{\mu}^A $) exhibits a trivial profile which leads to a vanishing torsion 2 form $ \tau_{\mu \nu}(=2\partial_{[\mu}\tau_{\nu]}) $. On the other hand, the cloak one form ($ \tau_{\mu}^A $) of the nonrelativistic background exhibits a nontrivial profile when flavour D6 branes are added into the picture which therefore leads to a non vanishing ($ \tau_{\mu \nu} \neq 0$) value for the torsion 2 form.}. With various examples at our disposal, we explicitly construct the NR sigma model and explore several properties of it among which the derivation of the T-duality rules are the most significant one.

We show that under transverse T-duality these NR sigma models are mapped into relativistic sigma models those are defined over general relativity backgrounds. We find out the associated Lorentzian metric and compute the central charge of the dual QFTs. Since these Lorentzian manifolds are related to the parent non-Lorentzian manifolds by means of a T-duality, therefore we claim that these are precisely the central charges associated with the quantum mechanical system dual to those non-Lorentzian manifolds.

Below, we interpret this NR limit as a ``quantum mechanical''/NR limit of the dual $ \mathcal{N}=2 $ SCFTs in $ 4d $. The detailed answer to this question is not known yet, however, we outline some steps along the way those might be be pursed as a possible future direction.
%%%%%%%%%%%%%%%%%%%%%%%%%%%%
\subsection{Couplings in the NR limit} 
In what follows, we try to gain some insights into the nature of the coupling constants in the NR limit of the $ \mathcal{N}=2 $ SCFTs. On the relativistic side of the story, these coupling constants are realised as the VEV of a real scalar field $ \langle \Phi_i \rangle $ which corresponds to the location of the $ i $th NS5 brane along the holographic axis. 

Typically, the Lagrangian of the dual QFT takes the form
\begin{eqnarray}
\mathcal{S}_{QFT}\sim \langle \Phi_{i+1} -\Phi_i  \rangle \int F_{mn}^2 \sim \frac{1}{g_{YM}^2} \int F_{mn}^2,
\end{eqnarray}
where $ m , n $ are the flat world-volume directions of the color D4 brane.

We start by considering the relativistic theory of color D4 brane probes on the ST background and thereby taking the NR limit following the scaling rules of (\ref{E2.50}). As a warm up, we first revisit the abelian T-dual example of \cite{Lozano:2016kum} and take its NR limit. 

To compute the coupling, the world-volume electromagnetic fields are turned on for the color D4 brane which is considered to be extended along the holographic $ \eta $ direction of the bulk and is fixed at a constant raial distance $ R $ of the $ AdS_5 $ factor. The remaining world-volume directions of the brane are along $ \mathbb{R}^{1,3} $ of the Minkowski ($\sim \frac{4R^2}{L^2} dx^2_{1,3} $) factor.

The relativistic action is given by \cite{Lozano:2016kum}
\begin{eqnarray}
S^{ATD}_{BIWZ}&=&-T_{D4}\left(\frac{4R^2}{L^2} \right)^2 \sqrt{\alpha'}\int d^5 x \left[ \sqrt{1- 4\pi^2 \alpha'^2 \left(\frac{L^2}{4R^2} \right)^2 F^2_{tx} }-1\right] \\
& \approx &2 \pi^2 \alpha'^{5/2} T_{D4}\int d^5x F^2_{tx}+\cdots ,
\end{eqnarray}
which is expanded considering the combination $ 4\pi^2 \alpha'^2  F^2_{tx}$ being small.

We now discuss the NR scaling that leads towards TSNC backgrounds. Under NR scaling, the world-volume factor of the D4 brane scales as
\begin{eqnarray}
d^5x = c^2 d^5 \tilde{x},
\end{eqnarray}
since the remaining world-volume directions are along the transverse directions of the brane. 

Together with this, we propose the following NR scaling\footnote{Here we consider the fact that under NR scaling (which leads to TSNC backgrounds) the tension of the Dp brane scales as $ T_{Dp}= \frac{\tilde{T}_{Dp}}{c^{p-1}}$.}
\begin{eqnarray}
T_{D4}=\frac{\tilde{T}_{D4}}{c^3}~;~\alpha' F_{tx} =  \frac{\alpha'_{NR}}{c^2}f_{\tilde{t}x},
\end{eqnarray}
which finally reveals
\begin{eqnarray}
S^{(NR)}_{ATD} =\frac{2 \pi^2}{c^4} \tilde{T}_{D4}\alpha'^{5/2}_{NR} \int d^5\tilde{x}f^2_{\tilde{t}x}.
\end{eqnarray}

A more involved example includes the case of NATD background for which the WZ action for D4 brane takes the following form \cite{Lozano:2016kum}
\begin{eqnarray}
S^{NATD}_{BIWZ} = -T_{D4}\int d^5x \frac{16 R^4 \eta}{\sqrt{\alpha'} L^3}\left[ \left(1+\frac{L^2}{\eta^2}\right)^{1/2} \sqrt{1- 4\pi^2 \alpha'^2 \left(\frac{L^2}{4R^2} \right)^2 F^2_{tx} }-1\right] .\nonumber\\
\end{eqnarray}

Using the NR scaling $ t =c \tilde{t} $, $ \eta =c \tilde{\eta} $, $ L= c^2 \ell $, $ R = c^2 \tilde{R} $ and $ \alpha'_{NR}=\frac{\alpha'}{c^2} $ we find
\begin{eqnarray}
\label{5.8}
S^{(NR)}_{NATD} \approx \frac{2 \pi^2 \ell^2}{c^2}  \alpha_{NR}^{'3/2}\tilde{T}_{D4}\int d^5 \tilde{x}f^2_{\tilde{t}x}=\frac{1}{g^2_{NR}}\int d^5 \tilde{x}f^2_{\tilde{t}x}.
\end{eqnarray}

The above calculation shows that the \emph{effective} coupling in the NR limit goes as $ g^2_{NR}\sim c^2 $. The $ \frac{1}{c^2} $ factor in the NR action (\ref{5.8}) is crucial and could be explained in two ways. However, in either ways, it is precisely an artefact of the NR scaling $ \eta_i \sim \langle \Phi_i\rangle \sim \frac{\eta_i}{c}  $ of NS5 branes along the holographic $ \eta $ axis.

Notice that, in the relativistic $ \mathcal{N}=2 $ Hanany-Witten set up, the color D4 branes are stretched between the NS5 branes (see Fig. \ref{hw}). As the locations of these NS5 branes are shifted (by a factor of $ 1/c $) in the large $ c\rightarrow \infty $ limit, the previous location of the centre of mass of the associated D4 world-volume direction is also shifted by a factor of $ 1/c $. The other way to interpret the $ \frac{1}{c^2} $ factor is to consider the effective coupling of the relativistic QFT that goes as, $ \frac{1}{g^2_4}\sim \frac{\Phi_{i+1}-\Phi_i}{g_s \sqrt{\alpha'}} $ \cite{Lozano:2016kum} which produces an additional factor of $ 1/c $ due to the presence of $ \sqrt{\alpha'} $ in the denominator.

To summarise, the above scaling (\ref{5.8}) of the effective coupling clearly shows that the separation $ \langle \Phi_{i+1} -\Phi_i  \rangle \sim \frac{1}{c} $ between NS5 branes goes to zero in the strict NR limit where all the NS5 branes are shifted towards the origin resulting in a metric singularity around\footnote{See (\ref{3.42}) and the discussions thereafter.} $ \eta \sim 0 $. In other words, this is the limit in which the coupling of the dual nonrelativistic quantum theory grows to infinity making the theory strongly coupled.
%%%%%%%%%%%%%%%%%%%%%%%%%%%%%%%%%
\subsection{Nonrelativistic limits of $ \mathcal{N}=2 $ SYM}
We conclude with an outline of obtaining the NR limit of $ \mathcal{N}=2 $ supersymmetric Lagrangian \cite{Seiberg:1994rs} using the null reduction procedure. $\mathcal{N}=2 $ supersymmetric models with gauge and matter multiplets can be constructed using $ \mathcal{N}=1 $ supersymmetric models.

\begin{figure}
\includegraphics[scale=.70]{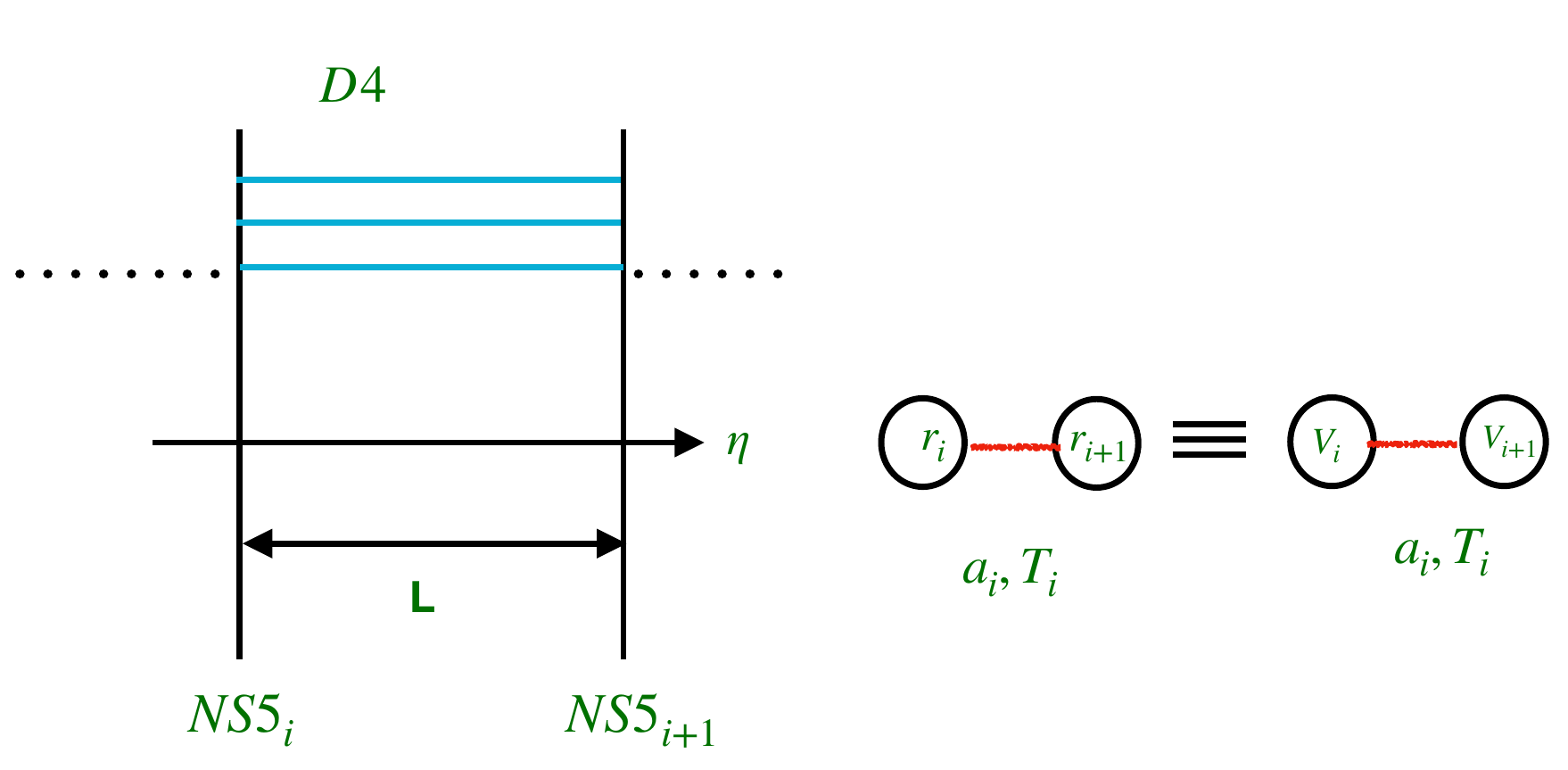}
  \caption{The above figure elaborates the Hanay-Witten brane set up of the $ \mathcal{N}=2 $ SCFTs and the associated quiver structure. The black circles essntially are the nodes those represent vector multiplets which transform in a particular adjoint representation of the gauge group $ U(r_i) $. These nodes are connected to each other thorugh (red) links which represent hypermultiplets of $ \mathcal{N}=2 $ theory. Together these nodes and links correspond to what is known as the $ \mathcal{N}=2 $ linear quiver. In the Hanany-Witten brane set up, these nodes are realised as the world-volume theory of $ r_i $ D4 branes stretched between the NS5$ _i $ and NS5$ _{i+1} $ branes. } \label{quiv}
\end{figure}

The field contents of $ \mathcal{N}=2 $ vector multiplet is built by putting together the scalar multiplet $ (A ,\psi) $ as well as the vector multiplet $ (A_{\mu}, \lambda) $ of $ \mathcal{N}=1 $ theory. To form an $ \mathcal{N}=2 $ vector multiplet, all these fields need to be transformed in the adjoint representation of the underlying gauge group $ U(r_i) $ where $ r_i $ corresponds to the rank of the gauge group. 

While constructing the quiver, these vector multiplets belonging to different color groups (of different rank) are denoted as ``nodes''. An $ \mathcal{N}=2 $ quiver is a collection of such nodes which are connected to each other through ``links'' which are basically constructed out of tensor multiplets ($ T_i $) and the bi-fundamental matter $ (h, \psi_{\dot{\alpha}})_i =a_i $ (Fig. \ref{quiv}).

\paragraph{Vector multiplets and its NR reduction.} Let us now focus on a particular vector multiplet $ V_i =(A_{\mu}, \lambda_{\alpha}, D)_i  $ (for a given $ i =1, \cdots , N-1$) of the quiver. The $\mathcal{N}=2 $ supersymmetric Lagrangian can be expressed as
\begin{eqnarray}
\label{5.9}
\mathcal{L}_{v}=\frac{1}{g^2}\text{Tr}\left( -\frac{F^2_{\mu  \nu}}{4}+\frac{g^2 \theta}{32 \pi^2}F_{\mu \nu}\tilde{F}^{\mu \nu}+(D_{\mu}A)^{\dagger}(D^{\mu}A)-\frac{1}{2}[A^{\dagger},A]^2 -i \lambda \sigma^{\mu}D_{\mu}\bar{\lambda}+\cdots\right), 
\end{eqnarray}
where we ignore spinors for simplicity. 

The $ \mathcal{N}=2 $ SYM is defined over the $ 4d $ Minkowski space labelled by the coordinates $ X^{\mu}~(\mu =0, \cdots , 3 )$. The space is endowed with the metric of the following form
\begin{eqnarray}
ds^2_4 = -(dx^0)^2 + (dx^{1})^2+\delta_{ij}dx^i dx^j ~;~i,j=2,3.
\end{eqnarray}

In what follows, we introduce light cone coordinates of the following form
\begin{eqnarray}
x^{\pm}=\frac{1}{\sqrt{2}}(x^0 \pm x^1),
\end{eqnarray}
which yields the Minkowski metric of the following form
\begin{eqnarray}
ds^2_4 =-2dx^{+}dx^{-}+\delta_{ij}dx^i dx^j .
\end{eqnarray}

Below, we estimate each of the individual terms in (\ref{5.9}) separately.
\begin{eqnarray}
\text{Tr} F^2_{\mu \nu}=\text{Tr} \mathfrak{f}^2_{ij}~;~\mathfrak{f}_{ij}=\partial_{[i}\mathfrak{a}_{j]}+[\mathfrak{a}_i ,\mathfrak{a}_j]~;~\text{Tr}F_{\mu \nu}\tilde{F}^{\mu \nu}=0,
\end{eqnarray}
\begin{eqnarray}
\text{Tr}(D_{\mu}A)^{\dagger}(D^{\mu}A)=\delta^{ij}\partial_i A^{\dagger}\partial_{j}A \text{Tr} \mathbb{I}-i \partial_k A^{\dagger}\mathfrak{a}^{m}_k T^m A \nonumber\\
+iA^{\dagger}(\mathfrak{a}^m_k T^m)^{\dagger}\partial_k A +A^{\dagger}(\mathfrak{a}^m_k T^m)^{\dagger}(\mathfrak{a}^n_k T^n)A,
\end{eqnarray}
\begin{eqnarray}
\text{Tr} \lambda \sigma^{\mu}D_{\mu}\bar{\lambda}=\lambda^m \sigma^+ \partial_+ \bar{\lambda}^m + \lambda^m \sigma^k \partial_k \bar{\lambda}^m \nonumber\\
+\lambda^m \sigma^+ f^{mnp}\mathfrak{a}^n_+ \bar{\lambda}^p +\lambda^m \sigma^k  f^{mnp}\mathfrak{a}^n_k \bar{\lambda}^p,
\end{eqnarray}
where we choose to work with the gauge, $ A_- =0 $.

Taking NR limit of the $ \mathcal{N}=2 $ quiver essentially corresponds to gluing several such multiplets together using ``links'' those might be realised as a NR limit of tensor as well as hypermultiplets. Below, we outline NR reduction of hypermultiplets those glue vector multiplets ($ V_i $) of different rank ($ r_i $).

\paragraph{Hypermultiplets.}  The Lagrangian of the hypermultiplets are typically of the form
\begin{eqnarray}
\mathcal{L}_h \sim | D_{\mu}\phi |^2 +i \bar{\psi}\slashed{D}\psi.
\end{eqnarray}

Let us focus on the bosonic part of the Lagrangian
\begin{eqnarray}
|D_{\mu}\phi |^2= |\partial_{\mu}\phi^{\alpha \beta} + g_1 A_{\mu}^{ab}\phi^{\alpha \beta}+g_2 A_{\mu}^{cd}\tilde{\phi}^{\tilde{\alpha}\tilde{\beta}}|^2,
\end{eqnarray}
where $ g_1 $ and $ g_2 $ are the couplings at two nodal points of the quiver. 

Here, $ \alpha =1, \cdots , N_i $ and $ \beta = 1, \cdots , N_{i+1} $ are the \emph{flavour} indices. Moreover, the indices $ b $ and $ \alpha $ are of equal weight. Clearly, the NR reduction would produce terms of the type $ (\partial_k \phi^{\alpha \beta})^2 $, $ g_1 g_2 \mathfrak{a}_{k}^{ab}\phi^{\alpha \beta}\mathfrak{a}_{k}^{cd}\tilde{\phi}^{\tilde{\alpha}\tilde{\beta}} , \cdots$ in the Lagrangian.

We conclude here by stating the fact, that at this point, it is not clear how to make sense of the above NR reduction in terms of the string theory results computed in this paper. This should be regarded as the ideal ground for future investigations, where one might be enthusiastic enough to compute various field theory observables using the NR reduction of $ \mathcal{N}=2 $ SCFTs and match them directly with the stringy counterpart.

%%%%%%%%%%%%%%%%%%%%%%%%%
%\appendix
%\section{Some title}

%%%%%%%%%%%%%%%%%%%%%%%%%%%
\acknowledgments
The author would like to thank Carlos Nunez for making several useful comments on the draft. 
The author is indebted to the authorities of IIT Roorkee for their
unconditional support towards researches in basic sciences. The author acknowledges The Royal Society, UK
for financial assistance and the Grant (No. SRG/2020/000088) received
from The Science and Engineering Research Board (SERB), India.
%%%%%%%%%%%%%%%%%%%%%%%%%%%%%%%%%%%%%%%%


\begin{thebibliography}{99}
\bibitem{Gomis:2000bd}
J.~Gomis and H.~Ooguri,
``Nonrelativistic closed string theory,''
J. Math. Phys. \textbf{42}, 3127-3151 (2001)
doi:10.1063/1.1372697
[arXiv:hep-th/0009181 [hep-th]].

\bibitem{Gomis:2005pg}
J.~Gomis, J.~Gomis and K.~Kamimura,
``Non-relativistic superstrings: A New soluble sector of AdS(5) x S**5,''
JHEP \textbf{12}, 024 (2005)
doi:10.1088/1126-6708/2005/12/024
[arXiv:hep-th/0507036 [hep-th]].

\bibitem{Bergshoeff:2018yvt}
E.~Bergshoeff, J.~Gomis and Z.~Yan,
``Nonrelativistic String Theory and T-Duality,''
JHEP \textbf{11}, 133 (2018)
doi:10.1007/JHEP11(2018)133
[arXiv:1806.06071 [hep-th]].

\bibitem{Bergshoeff:2015uaa}
E.~Bergshoeff, J.~Rosseel and T.~Zojer,
``Newton\textendash{}Cartan (super)gravity as a non-relativistic limit,''
Class. Quant. Grav. \textbf{32}, no.20, 205003 (2015)
doi:10.1088/0264-9381/32/20/205003
[arXiv:1505.02095 [hep-th]].

\bibitem{Bergshoeff:2019pij}
E.~A.~Bergshoeff, J.~Gomis, J.~Rosseel, C.~Simsek and Z.~Yan,
``String Theory and String Newton-Cartan Geometry,''
J. Phys. A \textbf{53}, no.1, 014001 (2020)
doi:10.1088/1751-8121/ab56e9
[arXiv:1907.10668 [hep-th]].

\bibitem{Harmark:2017rpg}
T.~Harmark, J.~Hartong and N.~A.~Obers,
``Nonrelativistic strings and limits of the AdS/CFT correspondence,''
Phys. Rev. D \textbf{96}, no.8, 086019 (2017)
doi:10.1103/PhysRevD.96.086019
[arXiv:1705.03535 [hep-th]].

\bibitem{Harmark:2018cdl}
T.~Harmark, J.~Hartong, L.~Menculini, N.~A.~Obers and Z.~Yan,
``Strings with Non-Relativistic Conformal Symmetry and Limits of the AdS/CFT Correspondence,''
JHEP \textbf{11}, 190 (2018)
doi:10.1007/JHEP11(2018)190
[arXiv:1810.05560 [hep-th]].

\bibitem{Gomis:2020izd}
J.~Gomis, Z.~Yan and M.~Yu,
``T-Duality in Nonrelativistic Open String Theory,''
JHEP \textbf{02}, 087 (2021)
doi:10.1007/JHEP02(2021)087
[arXiv:2008.05493 [hep-th]].

\bibitem{Hartong:2021ekg}
J.~Hartong and E.~Have,
``Nonrelativistic Expansion of Closed Bosonic Strings,''
Phys. Rev. Lett. \textbf{128}, no.2, 021602 (2022)
doi:10.1103/PhysRevLett.128.021602
[arXiv:2107.00023 [hep-th]].

\bibitem{Harmark:2019upf}
T.~Harmark, J.~Hartong, L.~Menculini, N.~A.~Obers and G.~Oling,
``Relating non-relativistic string theories,''
JHEP \textbf{11}, 071 (2019)
doi:10.1007/JHEP11(2019)071
[arXiv:1907.01663 [hep-th]].

\bibitem{Bidussi:2021ujm}
L.~Bidussi, T.~Harmark, J.~Hartong, N.~A.~Obers and G.~Oling,
``Torsional string Newton-Cartan geometry for non-relativistic strings,''
[arXiv:2107.00642 [hep-th]].

\bibitem{Yan:2021lbe}
Z.~Yan,
``Torsional deformation of nonrelativistic string theory,''
JHEP \textbf{09}, 035 (2021)
doi:10.1007/JHEP09(2021)035
[arXiv:2106.10021 [hep-th]].

\bibitem{Alvarez:1994wj}
E.~Alvarez, L.~Alvarez-Gaume and Y.~Lozano,
``A Canonical approach to duality transformations,''
Phys. Lett. B \textbf{336}, 183-189 (1994)
doi:10.1016/0370-2693(94)00982-1
[arXiv:hep-th/9406206 [hep-th]].

\bibitem{Alvarez:1994dn}
E.~Alvarez, L.~Alvarez-Gaume and Y.~Lozano,
``An Introduction to T duality in string theory,''
Nucl. Phys. B Proc. Suppl. \textbf{41}, 1-20 (1995)
doi:10.1016/0920-5632(95)00429-D
[arXiv:hep-th/9410237 [hep-th]].

\bibitem{Gaiotto:2009we}
D.~Gaiotto,
``N=2 dualities,''
JHEP \textbf{08}, 034 (2012)
doi:10.1007/JHEP08(2012)034
[arXiv:0904.2715 [hep-th]].

\bibitem{Gaiotto:2009gz}
D.~Gaiotto and J.~Maldacena,
``The Gravity duals of N=2 superconformal field theories,''
JHEP \textbf{10}, 189 (2012)
doi:10.1007/JHEP10(2012)189
[arXiv:0904.4466 [hep-th]].

\bibitem{Sfetsos:2010uq}
K.~Sfetsos and D.~C.~Thompson,
``On non-abelian T-dual geometries with Ramond fluxes,''
Nucl. Phys. B \textbf{846}, 21-42 (2011)
doi:10.1016/j.nuclphysb.2010.12.013
[arXiv:1012.1320 [hep-th]].

\bibitem{Maldacena:2000mw}
J.~M.~Maldacena and C.~Nunez,
``Supergravity description of field theories on curved manifolds and a no go theorem,''
Int. J. Mod. Phys. A \textbf{16}, 822-855 (2001)
doi:10.1142/S0217751X01003937
[arXiv:hep-th/0007018 [hep-th]].

\bibitem{ReidEdwards:2010qs}
R.~A.~Reid-Edwards and B.~Stefanski, jr.,
``On Type IIA geometries dual to N = 2 SCFTs,''
Nucl. Phys. B \textbf{849}, 549-572 (2011)
doi:10.1016/j.nuclphysb.2011.04.002
[arXiv:1011.0216 [hep-th]].

\bibitem{Aharony:2012tz}
O.~Aharony, L.~Berdichevsky and M.~Berkooz,
``4d N=2 superconformal linear quivers with type IIA duals,''
JHEP \textbf{08}, 131 (2012)
doi:10.1007/JHEP08(2012)131
[arXiv:1206.5916 [hep-th]].

\bibitem{Lozano:2016kum}
Y.~Lozano and C.~N\'u\~nez,
``Field theory aspects of non-Abelian T-duality and $ \mathcal{N}  =$ 2 linear quivers,''
JHEP \textbf{05}, 107 (2016)
doi:10.1007/JHEP05(2016)107
[arXiv:1603.04440 [hep-th]].

\bibitem{Nunez:2019gbg}
C.~N\'u\~nez, D.~Roychowdhury, S.~Speziali and S.~Zacar\'\i{}as,
``Holographic aspects of four dimensional ${\cal N }=2$ SCFTs and their marginal deformations,''
Nucl. Phys. B \textbf{943}, 114617 (2019)
doi:10.1016/j.nuclphysb.2019.114617
[arXiv:1901.02888 [hep-th]].

\bibitem{Lozano:2017ole}
Y.~Lozano, C.~Nunez and S.~Zacarias,
``BMN Vacua, Superstars and Non-Abelian T-duality,''
JHEP \textbf{09}, 008 (2017)
doi:10.1007/JHEP09(2017)008
[arXiv:1703.00417 [hep-th]].

\bibitem{Berenstein:2002jq}
D.~E.~Berenstein, J.~M.~Maldacena and H.~S.~Nastase,
``Strings in flat space and pp waves from N=4 superYang-Mills,''
JHEP \textbf{04}, 013 (2002)
doi:10.1088/1126-6708/2002/04/013
[arXiv:hep-th/0202021 [hep-th]].

\bibitem{Henneaux:2008nr}
M.~Henneaux, E.~Jamsin, A.~Kleinschmidt and D.~Persson,
``On the E10/Massive Type IIA Supergravity Correspondence,''
Phys. Rev. D \textbf{79}, 045008 (2009)
doi:10.1103/PhysRevD.79.045008
[arXiv:0811.4358 [hep-th]].

\bibitem{Bergshoeff:2021bmc}
E.~A.~Bergshoeff, J.~Lahnsteiner, L.~Romano, J.~Rosseel and C.~\c{S}im\c{s}ek,
``A non-relativistic limit of NS-NS gravity,''
JHEP \textbf{06}, 021 (2021)
doi:10.1007/JHEP06(2021)021
[arXiv:2102.06974 [hep-th]].

\bibitem{kova}J.J. Kovacic, An algorithm for solving second order linear homogeneous differential equations,
J. Symbolic Comput. 2 (1986) 3.x

\bibitem{Basu:2011fw}
P.~Basu and L.~A.~Pando Zayas,
``Analytic Non-integrability in String Theory,''
Phys. Rev. D \textbf{84}, 046006 (2011)
doi:10.1103/PhysRevD.84.046006
[arXiv:1105.2540 [hep-th]].

\bibitem{Stepanchuk:2012xi}
A.~Stepanchuk and A.~A.~Tseytlin,
``On (non)integrability of classical strings in p-brane backgrounds,''
J. Phys. A \textbf{46}, 125401 (2013)
doi:10.1088/1751-8113/46/12/125401
[arXiv:1211.3727 [hep-th]].

\bibitem{Nunez:2018qcj}
C.~Nunez, D.~Roychowdhury and D.~C.~Thompson,
``Integrability and non-integrability in $ \mathcal{N}=2 $ SCFTs and their holographic backgrounds,''
JHEP \textbf{07}, 044 (2018)
doi:10.1007/JHEP07(2018)044
[arXiv:1804.08621 [hep-th]].

\bibitem{Roychowdhury:2019olt}
D.~Roychowdhury,
``Nonrelativistic pulsating strings,''
JHEP \textbf{09}, 002 (2019)
doi:10.1007/JHEP09(2019)002
[arXiv:1907.00584 [hep-th]].

\bibitem{Macpherson:2014eza}
N.~T.~Macpherson, C.~N\'u\~nez, L.~A.~Pando Zayas, V.~G.~J.~Rodgers and C.~A.~Whiting,
``Type IIB supergravity solutions with AdS$_{5}$ from Abelian and non-Abelian T dualities,''
JHEP \textbf{02}, 040 (2015)
doi:10.1007/JHEP02(2015)040
[arXiv:1410.2650 [hep-th]].

\bibitem{Seiberg:1994rs}
N.~Seiberg and E.~Witten,
``Electric - magnetic duality, monopole condensation, and confinement in N=2 supersymmetric Yang-Mills theory,''
Nucl. Phys. B \textbf{426}, 19-52 (1994)
[erratum: Nucl. Phys. B \textbf{430}, 485-486 (1994)]
doi:10.1016/0550-3213(94)90124-4
[arXiv:hep-th/9407087 [hep-th]].

\end{thebibliography}
\end{document}